\documentclass[]{interact}
\usepackage{hyperref}
\usepackage{epstopdf}
\usepackage{algpseudocode}
\usepackage[caption=false]{subfig}

\usepackage[numbers,sort&compress]{natbib}
\usepackage{color, colortbl}
\bibpunct[, ]{[}{]}{,}{n}{,}{,}
\renewcommand\bibfont{\fontsize{10}{12}\selectfont}

\theoremstyle{plain}
\newtheorem{theorem}{Theorem}[section]

\newtheorem{corollary}[theorem]{Corollary}
\newtheorem{proposition}[theorem]{Proposition}

\theoremstyle{definition}

\newtheorem{example}[theorem]{Example}

\theoremstyle{remark}

\definecolor{LightCyan}{rgb}{0.88,1,1}

\begin{document}

\title{\textbf{Developing flexible classes of distributions to account for both skewness and bimodality}}
\author{Jamil Ownuk$^*$\footnote{$^*$Corresponding author, Email: jamilownuk@yahoo.com}, Ahmad Nezakati \& Hossein Baghishani \\
Department of Statistics, Faculty of Mathematical Sciences, Shahrood University of Technology, Shahrood, Iran}

\maketitle
\hspace*{-6mm}
\textbf{Abstract:} \\ \ \\
We develop two novel approaches for constructing skewed and bimodal flexible distributions that can effectively generalize classical symmetric distributions. We illustrate the application of introduced techniques by extending normal, student-t, and Laplace distributions. We also study the properties of the newly constructed distributions. The method of maximum likelihood is proposed for estimating the model parameters. Furthermore, the application of new distributions is represented using real-life data.
\section{Introduction}
In recent years, many statisticians have been introduced generalizations for statistical distributions. These extensions of distributions to the new one give more flexibility to the original probability density functions in practice. Some of these new developments are defined as follow.

\begin{description}
\item[a)] 
Azzalini (1985) introduced a skew family of densities as $ 2F(\lambda x) g(x) $, for any real $\lambda$, in which $g$ is a density function symmetric about 0, and $F$ an absolutely continuous distribution function such that its first derivative is symmetric about 0. For an example of this family, we can mention skew-normal (SN; Azzalini, 1985) and skew student-t (SSt; Arellano-Valle and Azzalini, 2013).
\item[b)]
Skew-Symmetric family of distributions with probability density function $ 2 \pi(x) g(x) $ was proposed by Arnold and Lin (2004) where $ \pi $ is a Lebesgue measurable function satisfying $ 0 \leq \pi(x) \leq 1 $ and $ \pi(x) + \pi(-x) = 1 $, and $g$ is a density function symmetric about 0. 
\item[c)]
Exponentiated type distributions due to Gupta et al. (1998) has the form 
 \begin{equation}
  \alpha \left( 1 - G(x) \right) ^ {\alpha - 1} g(x)  \nonumber
 \end{equation}
 Where $ g $ is any density function and $ G $ corresponding cumulative distribution function and, $\alpha > 0$ is a shape parameter.
\item[d)] 
Beta G distributions introduced by Eugene et al. (2002) has density function
  \begin{equation}
    \frac{1}{B(a,b)} G(x) ^ {a-1} \left( 1 - G(x) \right) ^ {b-1} g(x)   \nonumber
   \end{equation}
Where $ g $ is any density function, $ G $ its CDF, $a > 0, b > 0$ are shape parameters and $ B(a,b) $ is beta function. Beta Normal (BN) Eugene et al. (2002) distribution is an example of these family of distributions.
\item[e)]
The density function of odd log-logistic (OLL) family of distributions worked out by Gleaton and Lynch (2006) is given by 
\begin{equation}
        \frac{ \alpha G(x)^{\alpha - 1} \left( 1 - G(x) \right) ^ {\alpha - 1 } }{\left[ G(x)^{\alpha } + \left( 1 - G(x) \right) ^ {\alpha } \right] ^ 2} g(x)   \nonumber
       \end{equation}
where $\alpha > 0$ is shape parameter. The odd log-logistic Normal (OLLN) Duarte et al. (2018) is an example of the OLL family of distributions.
The density function of generalized odd log-logistic (GOLL) family of distributions studied by Cordeiroa et al. (2016) is 
    \begin{equation}
        \frac{ \alpha \theta  G(x)^{\alpha \theta - 1} \left( 1 - G(x) ^ \theta \right) ^ {\alpha - 1 } }{\left[ G(x)^{\alpha \theta} + \left( 1 - G(x) ^ \theta \right) ^ {\alpha } \right] ^ 2} g(x)   \nonumber
       \end{equation}
where $\alpha > 0$ and $\theta > 0$ are shape parameters.       
\end{description} 
In addition to the families mentioned above, many families of distributions were available in lectures, for examples exponentiated half-logistic introduced by Cordeiro et al. (2014), Kumaraswamy G distributions due to Cordeiro and Castro (2011), Weibull G distributions worked out by Alzaatreh et al. (2013b), Nadarajah, Cancho, and Ortega (2013) presented geometric exponential Poisson G distributions, truncated-exponential skew-symmetric G distributions mentioned by Nadarajah, Nassiri, and Mohammadpour (2014), Ristic and Nadarajah (2014) established exponentiated exponential Poisson G distributions, Bolfarine et al. (2018) constructed Bimodal symmetric-asymmetric power-normal (BAPN)
families of distributions and power log-Dagum (PLD) distribution worked out by Bakouch et al. (2019).   

The idea of this paper is applying new theorems to extend univariate distributions to new density functions which are skew and bimodal. We give some example to illustrate how to use these theorems. Also, we develop Normal, Student-t and Laplace distribution to the significant flexible skew and bimodal data.  

In section 2, we discuss constructing bimodal-unimodal distributions, and we will present some example. In section 3, we will obtain mathematical properties of some new distributions. In section 4, with three real data sets, we will show that the new distributions have more flexibility and more suitable.  
   
\section{Constructing bimodal and skewed families}
\subsection{The first approach}
\begin{figure}  [h!]
\centering 
\subfloat{
\resizebox*{6cm}{!}{\includegraphics{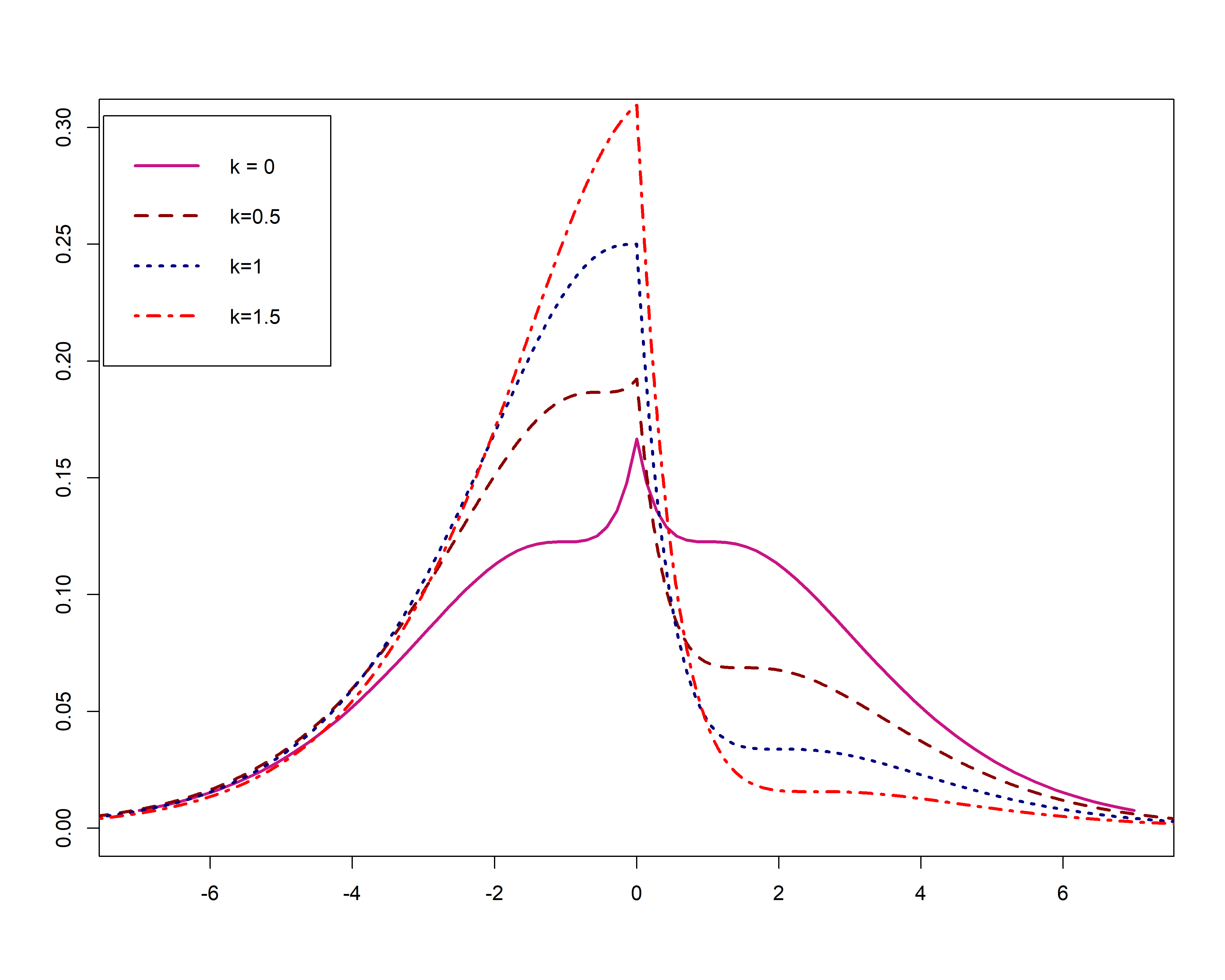}}}\hspace{5pt}
\subfloat{
\resizebox*{6cm}{!}{\includegraphics{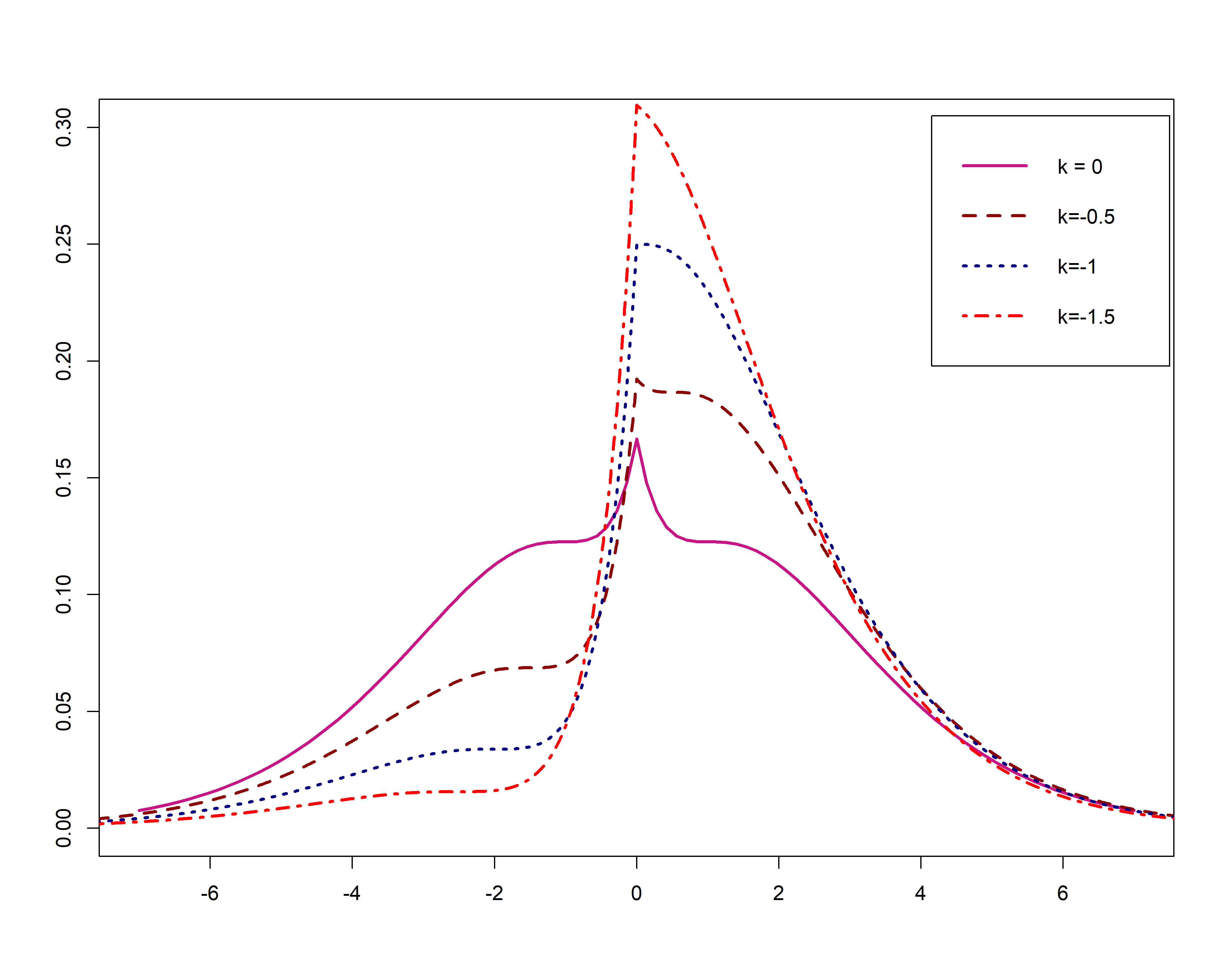}}}
\caption{Shapes of density function \eqref{eq:1} for $ \mu = 0 $, $ \sigma = 1 $ and some value of $k$.} \label{fig:1}
\end{figure}

Let us give you the idea of this approach with a simple example. One way for generalizing distributions to flexible one is product two pdf or function of it. for example, let $ g_1(x) $ is standard Laplace distribution, and $ g_2(x-k) $ is Cauchy distribution with location parameter $ k $ and scale parameter $1$, then
\begin{equation} \label{eq:1}
f(x) = \frac{1}{\pi \left( 3 + k ^ 2 \right)} \frac{g_1(x)}{g_2(x-k)}
\end{equation}
 is flexible density function. We call this new pdf as Laplace Cauchy (LC) distribution. LC distribution is symmetric when $ k = 0 $ and skew otherwise. It is more flexible and not complicated. Moments of LC distribution are given by
\begin{equation}
E\left( X^r \right) = \frac{2}{3 + k^2} \left\{ k ^ 2 E\left( X_{L} ^ {r}  \right) - 2 k E\left( X_{L} ^ {r+1}  \right) + E\left( X_{L} ^ {r+2}  \right)  \right\}  \nonumber
\end{equation}
where $ X_L $ is the standard Laplace random variable. Shapes of density function \eqref{eq:1} for some value of $ k $ are shown in Figure \eqref{fig:1}. 

The following theorem shows how to construct symmetric bimodal distributions.
\begin{theorem} \label{th:2.1}
 Let $g(\cdot)$ is a symmetric probability density function about $c$ with support $ \mathbb{R} $ and $w(\cdot)$ is a strictly positive convex symmetric function about $b$ as well and $E_{g}[w(X)]<\infty$. Hence, for $ b=c $ the density function 
 \begin{equation} \label{eq:2}
 f(x) = \frac{w(x)}{E_{g}[w(X)]} g(x)
 \end{equation} 
 \begin{description}
 \item[i)]
 symmetric about $ c $.
 \item[ii)] a bimodal density function
 \item[iii)] the sum of its two modes is equal to $2 c $. 
 \end{description}
\end{theorem}

\begin{example} \label{ex:2.2}
Let $ X \sim N(0,1) $ and $ w(x) = \exp\left( k \left| x \right|  \right)  $ , then
\begin{eqnarray} 
f_{X}(x) = \frac{\exp\left( k \left| x \right|  \right) }{E_{\phi} [\exp\left( k \left| X \right|  \right) ]} \phi(x)   \nonumber
\end{eqnarray}
where $ w(x) $ for $ k>0 $ is convex and $ f_{X}(x)$ is symmetric about zero, and its modes are $ \pm k $.
\end{example}

Theorem \eqref{th:2.1} shows that if $ w(\cdot) $ in \eqref{eq:2} convex then for some value of parameters we will have skew bimodal distributions for $ b \neq c $. we can conclude that for concave $ w(\cdot) $, \eqref{eq:2} will be unimodal distribution and symmetric for $ b = c $ and skew for $ b \neq c $.

To use \eqref{eq:2} for constructing flexible distributions, we have to obtain normalizing constant $E_{g}\left( w(X)\right) $, which could be troublesome. The following theorem provides the conditions under which there is no difficulty to calculate the normalizing constant. 
\begin{theorem} \label{th:2.3}
if $ g(x) $ is a pdf such that strictly positive and symmetric about zero, with cdf $ G_{X}$, then 
\begin{equation} \label{eq:3}
f_{X}(x) = \frac{k}{2 \left( e ^ k - e ^{\frac{k}{2}} \right) }  g(x) e ^ {k G \left( |x| \right) }
\end{equation}
 and
\begin{equation} \label{eq:4}
 \frac{k+1}{2 \left( 1 - \frac{1}{2^{k+1}} \right) } g(x) G\left( |x| \right) ^ k
\end{equation}
are symmetric density functions and they are bimodal for $ k > 0 $.
\end{theorem}

\begin{figure} [h!]
\centering
\subfloat[]{
\resizebox*{6cm}{!}{\includegraphics{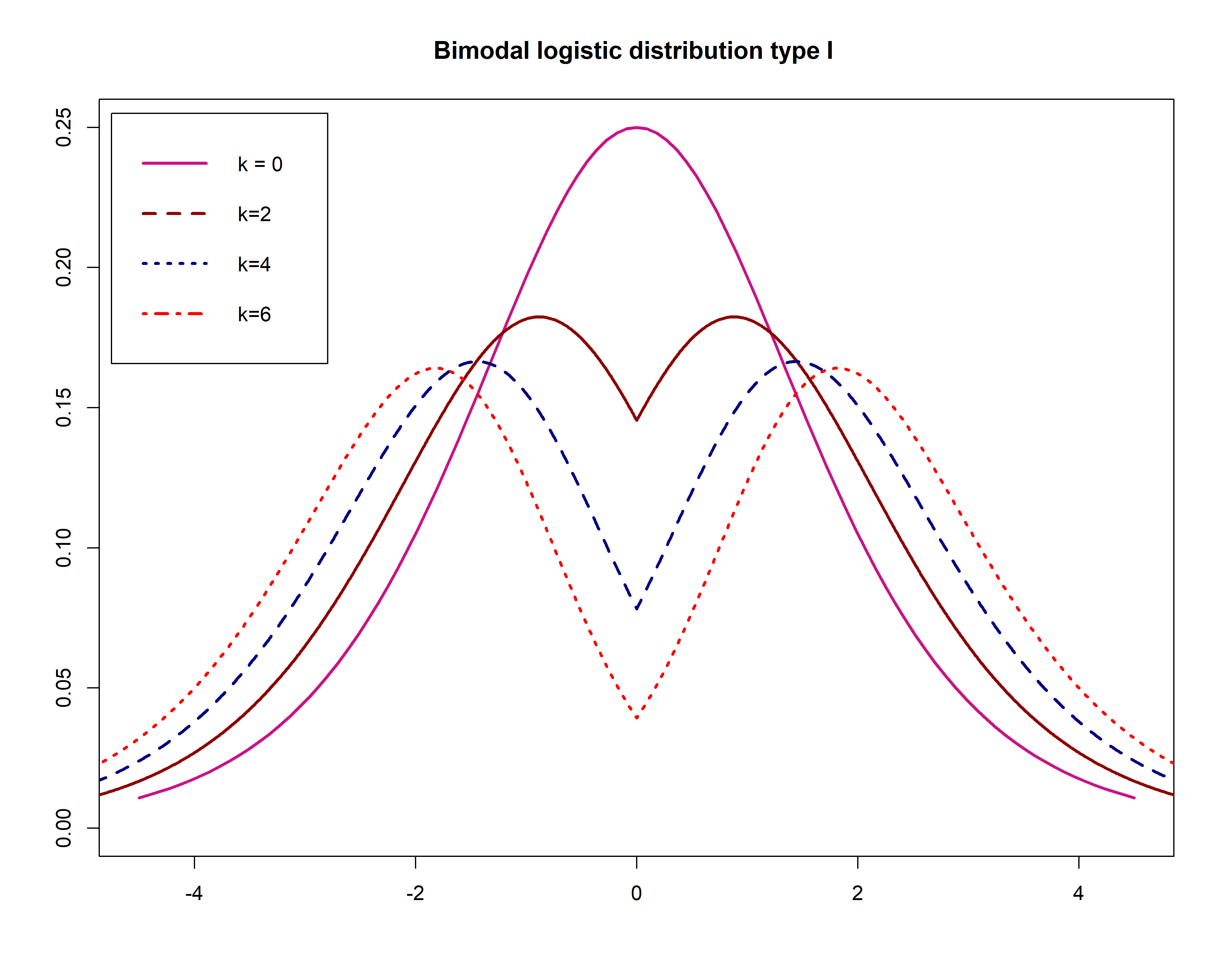}}}\hspace{5pt}
\subfloat[]{
\resizebox*{6cm}{!}{\includegraphics{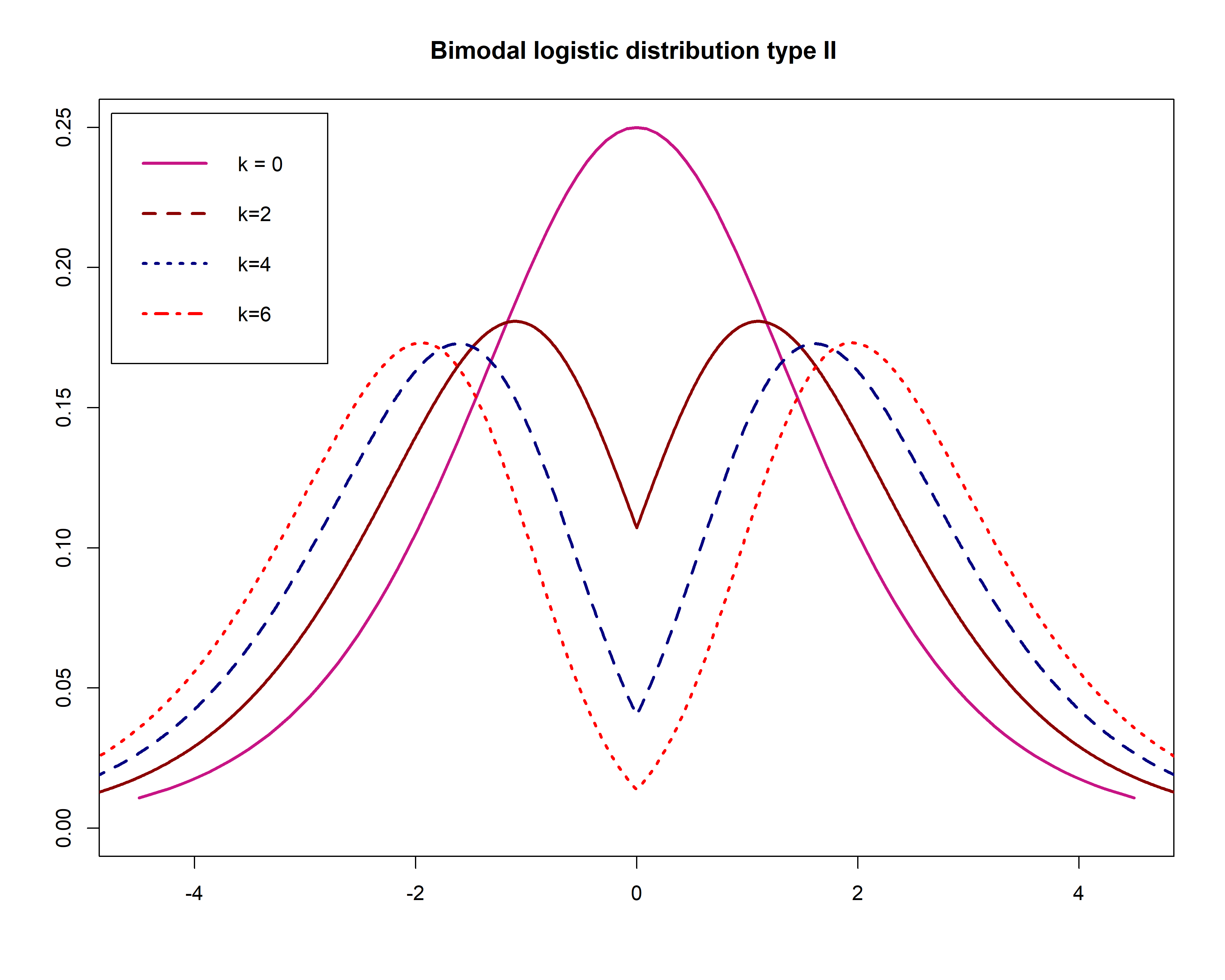}}}
\caption{Shapes of density function \eqref{eq:5} part (a) and \eqref{eq:6} part (b) for some value of $k$.} \label{fig:2}
\end{figure}

\begin{example} \label{ex:2.4}
Let $ X $ follow the standard logistic distribution. Density functions of $ f_X(x) $ in theorem \eqref{th:2.3} are given by
\begin{equation} \label{eq:5}
f_{X}(x) = \frac{k}{2 \left( e ^ k - e ^{\frac{k}{2}} \right) }  \left\{\frac{e^{-x}}{\left( 1 + e^{-x}\right) ^2}\right\} e^{\left\{\frac{k}{1 + e^{-|x|}}\right\}}
\end{equation}
 and
\begin{equation} \label{eq:6}
 f_{X}(x) = \frac{k+1}{2 \left( 1 - \frac{1}{2^{k+1}} \right) } \left\{\frac{e^{-x}}{\left( 1 + e^{-x}\right) ^2}\right\} \left\{\frac{1}{1 + e^{-|x|}}\right\} ^ k
\end{equation}
If random variable $X$ has pdf \eqref{eq:5} and \eqref{eq:6}, we say $X$ has Bimodal Logistic type I and type II distribution respectively. Shapes of density function \eqref{eq:5} and \eqref{eq:6} for some value of $k$ are shown in Figure \eqref{fig:2}.
\end{example} 

\subsection{The second approach}

Let $X$  has pdf $ g(x) $, and $h(\cdot)$ is  any positive, measurable function that $ \int_{-\infty}^{\infty} g\left( h(w) \right) dw < \infty $ then the new family of distributions is given by
\begin{eqnarray} \label{eq:7}
f_{X}(x) = \frac{g\left( h(x) \right) }{\int_{-\infty}^{\infty} g\left( h(w)  \right) dw } \cdot
\end{eqnarray}
The following theorem shows how to construct symmetric bimodal distributions from any unimodal distribution with support $\mathbb{R}$.
\begin{theorem} \label{th:2.5}
 if $ g(x) $ has a mode in $ k $ and $h(\cdot)$ are symmetric about $d$ and strictly convex function then pdf \eqref{eq:2} are
 \begin{description}
  \item[i)] 
  if $ k = 0 $, unimodal with the mode in $d$.
  \item[ii)] if $ k \ne 0 $, bimodal.
  \item[iii)] the mode(s) is (are) solution of $ h(x) = k $.
  \end{description}
\end{theorem}

\begin{example} \label{ex:2.6}
Let $ X \sim C(k,1) $ and $ h(x) = |x|  $ , then
\begin{eqnarray} \label{eq:8}
f_{X}(x) =\frac{1}{2G(k)} \frac{1}{\pi} \frac{1}{1+\left( |x|-k\right) ^2}  
\end{eqnarray} 
Where $ G(.) $ is cdf $ C(0,1) $, $ f_{X}(x)$ is symmetric about zero, and its modes are $ \pm k $.
\end{example}

\begin{example} \label{ex:2.7}
Let $ X $ follow hyperbolic secant distribution (Fisher (1921)) with location $k$ and scale 1 and $ h(x) = |x|  $ , then
\begin{eqnarray}  \label{eq:9}
f_{X}(x) =\frac{1}{2G(k)} \frac{2}{\pi} \frac{1}{e^{\left( |x|-k\right) } + e ^ {-\left( |x|-k\right)} }  
\end{eqnarray}
where $ G(.) $ is cdf $ HS(0,1) $, $ f_{X}(x)$ is symmetric about zero and its modes are $ \pm k $.
\end{example}

\begin{figure} [h!]
\centering
\subfloat[]{
\resizebox*{6cm}{!}{\includegraphics{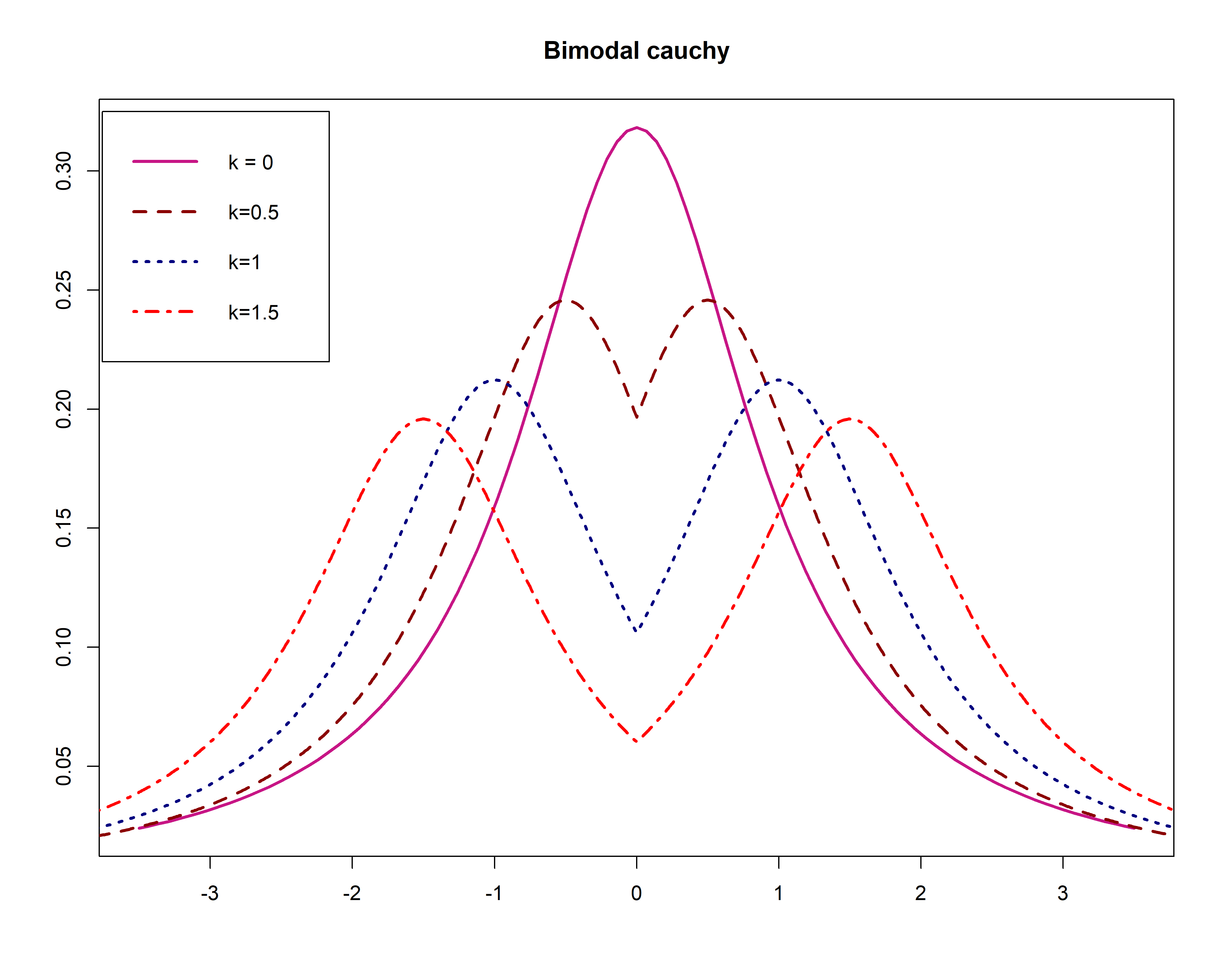}}}\hspace{5pt}
\subfloat[]{
\resizebox*{6cm}{!}{\includegraphics{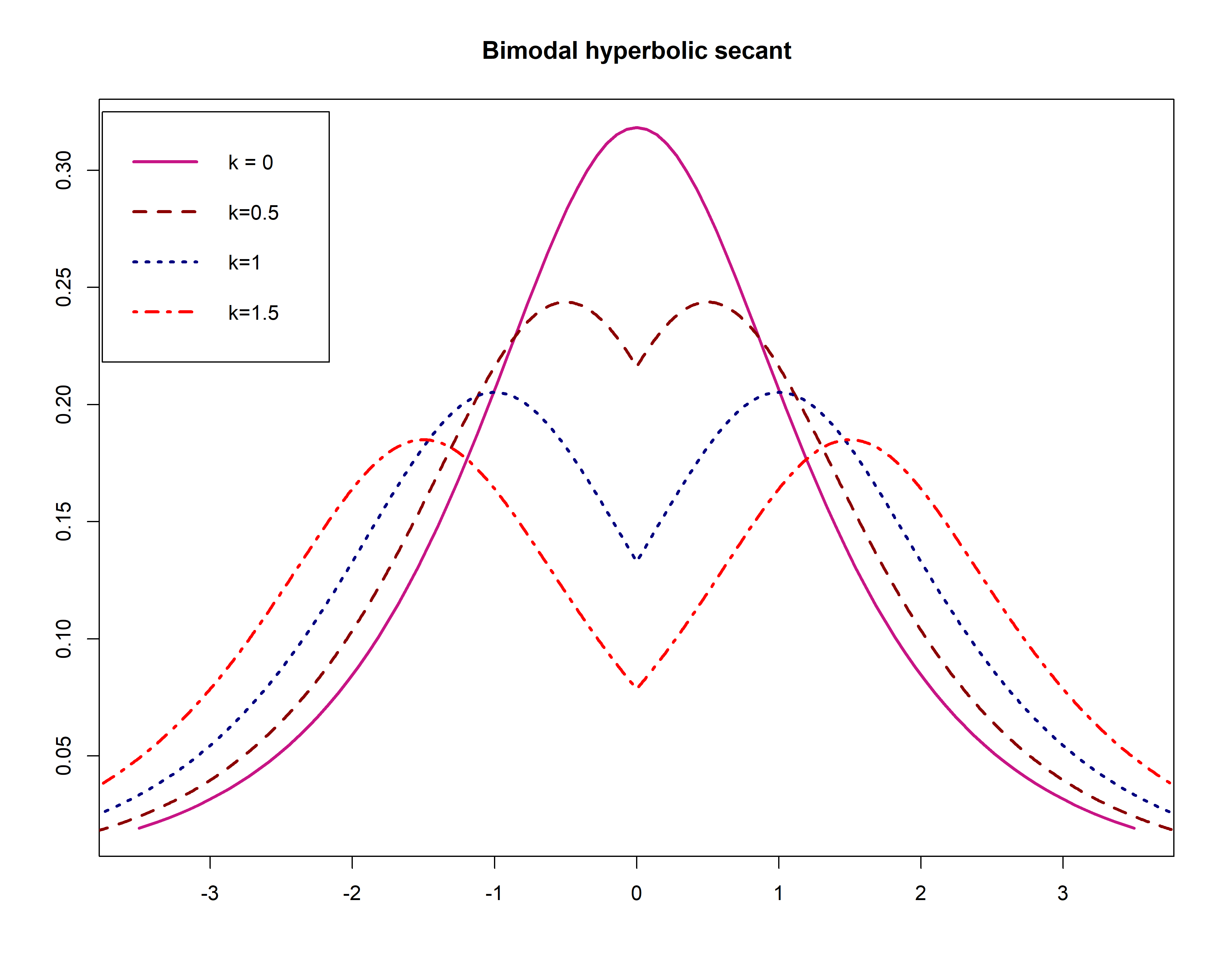}}}
\caption{Shapes of density function \eqref{eq:8} part (a) and \eqref{eq:9} part (b) for some value of $k$.} \label{fig:3}
\end{figure}

Shapes of density function \eqref{eq:8} and \eqref{eq:9} for some value of $k$ are shown in Figure \eqref{fig:3}.

The following theorem shows how to construct symmetric bimodal density function without difficulty to calculate normalizing constant. 
\begin{theorem} \label{th:2.8}
If $ g(x) $ is a pdf such that strictly positive and symmetric about $ k \ne 0 $, with cdf $ G_{X}$, then $ f_{X}(x) = \frac{1}{2}\frac{g\left( |x| \right) }{G_{(X-k)} (k) } $ is symmetric bimodal density function.
\end{theorem}

We can apply the idea of Azzalini type family of distributions to construct skewed density function for results of theorem \eqref{th:2.8}. The following corollary shows how we can use this idea.  
\begin{corollary}
With notation in theorem \eqref{th:2.8}, $ f_{X}(x) = \frac{g\left( |x| \right) }{G_{(X-k)} (k) } F\left(\lambda x \right) $ is a density function. Where $F\left(\cdot \right)$ is absolutely continuous distribution function such that its first derivative is symmetric about 0.
\end{corollary}

\begin{figure} [h!]
\centering
\subfloat{
\resizebox*{6cm}{!}{\includegraphics{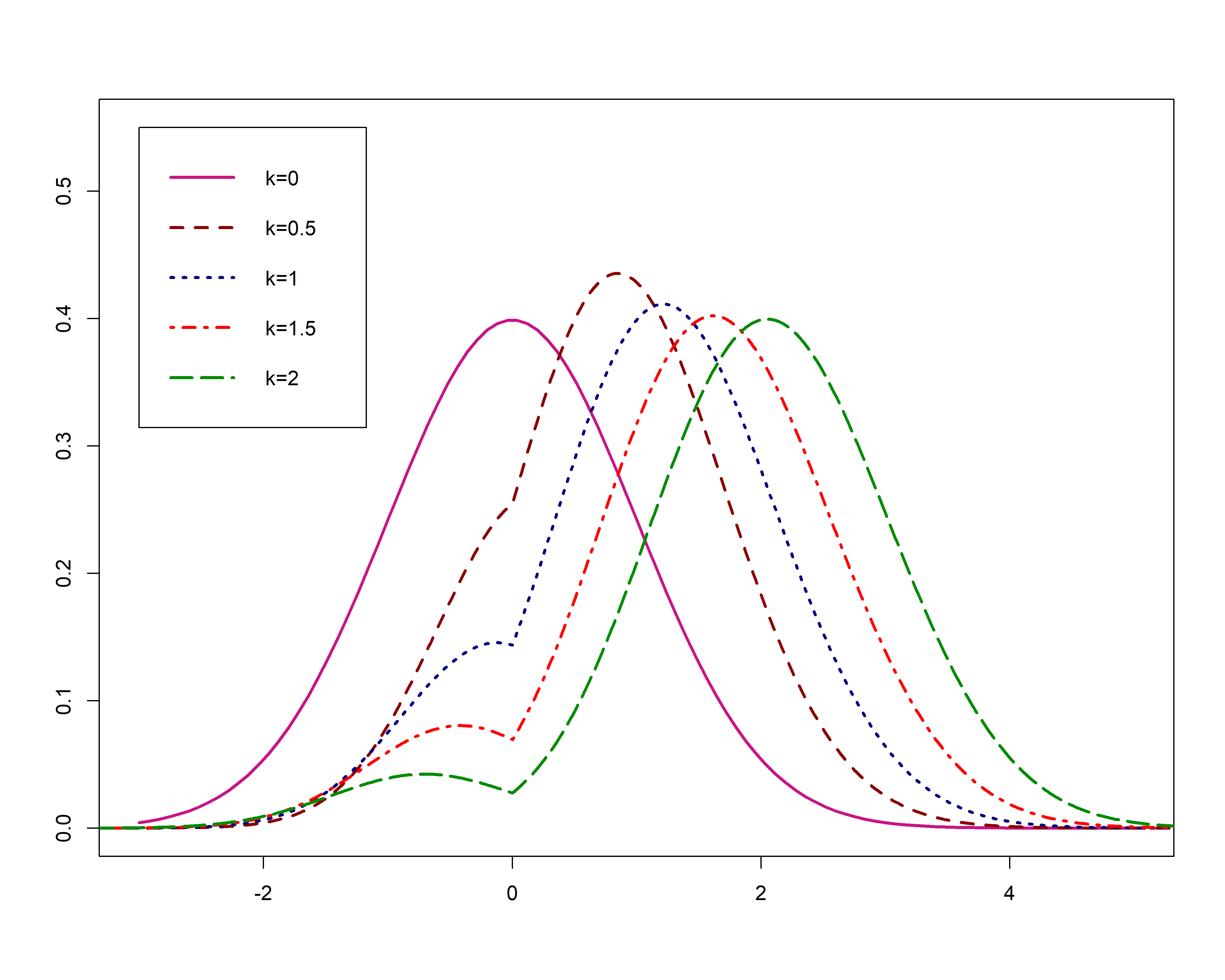}}}\hspace{5pt}
\subfloat{
\resizebox*{6cm}{!}{\includegraphics{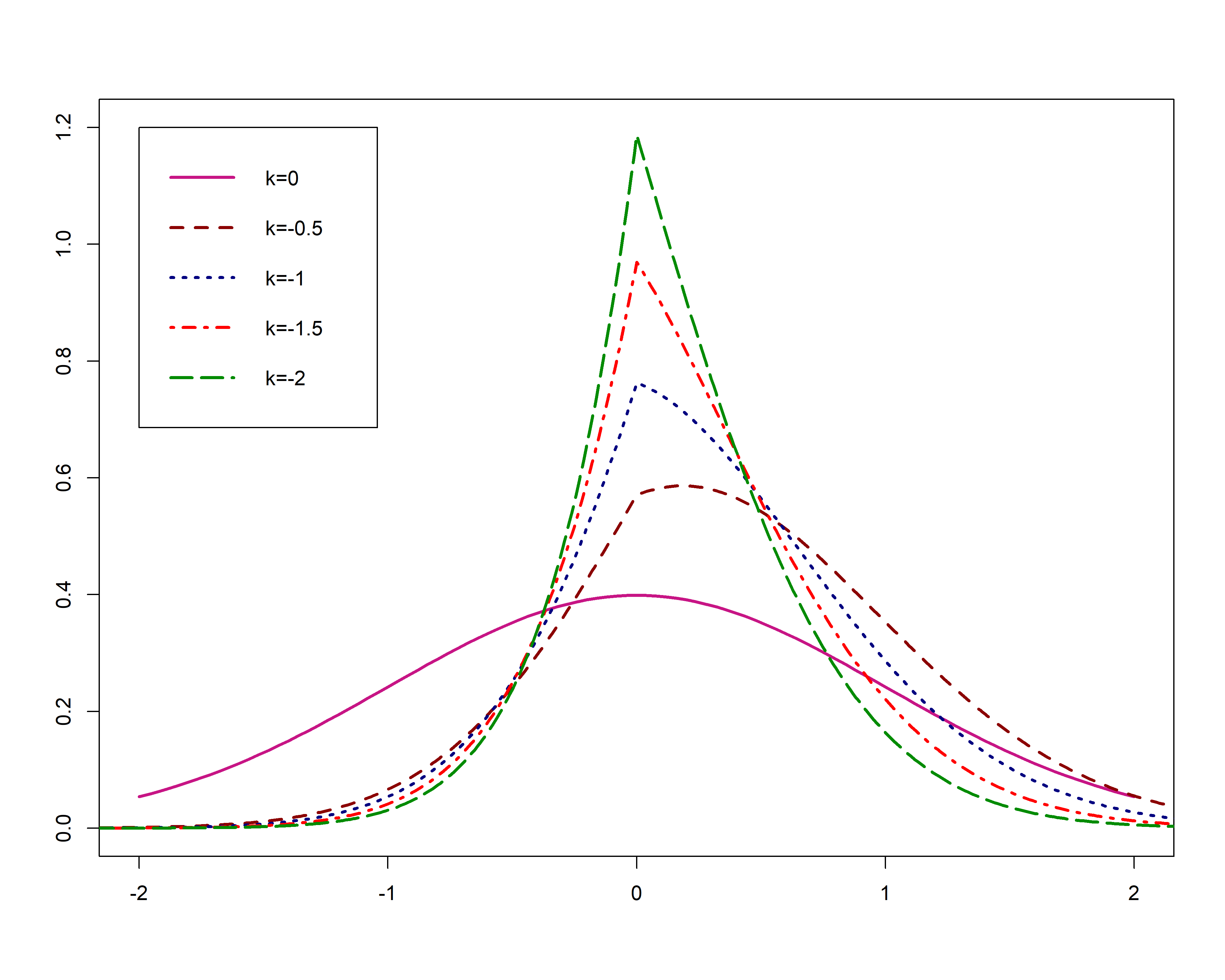}}}
\caption{Shapes of density function \eqref{eq:10} for some value of $k$.} \label{fig:4}
\end{figure}

\begin{example} 
Let $ X \sim N(k,1) $, then
\begin{eqnarray}  \label{eq:10}
f_{X}(x) = \phi\left( \left| x \right|-k\right) \frac{\Phi(\lambda x )}{\Phi\left( k \right) } \cdot
\end{eqnarray}

\end{example}
Shapes of density function \eqref{eq:10} for some value of parameters are shown in Figure \eqref{fig:4}.

\section{Developing Flexible Bimodal Distributions}
Normal, Student-t and Laplace distributions are more known and applicable models in symmetric data. Also, we can suggest these distributions as thick-tail, heavy-tail and semi-heavy-tail models. There are some generalizations of these famous models in literature which are applicable in one situation. For example, Skew Normal, Skew Student-t and Skew Laplace are usable for skewed data. In this section, with using theorems in the previous section, we construct new generalizations of Normal, Skew-t and Laplace distributions which have bimodal, unimodal, symmetric and asymmetric density functions. Advantage of this new generalizations are 
\begin{itemize}
\item Applicable for a large class of data as symmetric, skewed, bimodal data. 
\item Having Normal or Student-t or Laplace distributions as sub-models.
\item Having simple density functions so it is easy to work with them.
\item Some of these extensions have stochastic representation so we can easily generate data.
\end{itemize}     
\subsection{Bimodal-Unimodal Normal Distribution}     
\begin{figure} [h!]
\centering
\subfloat[]{
\resizebox*{6cm}{!}{\includegraphics{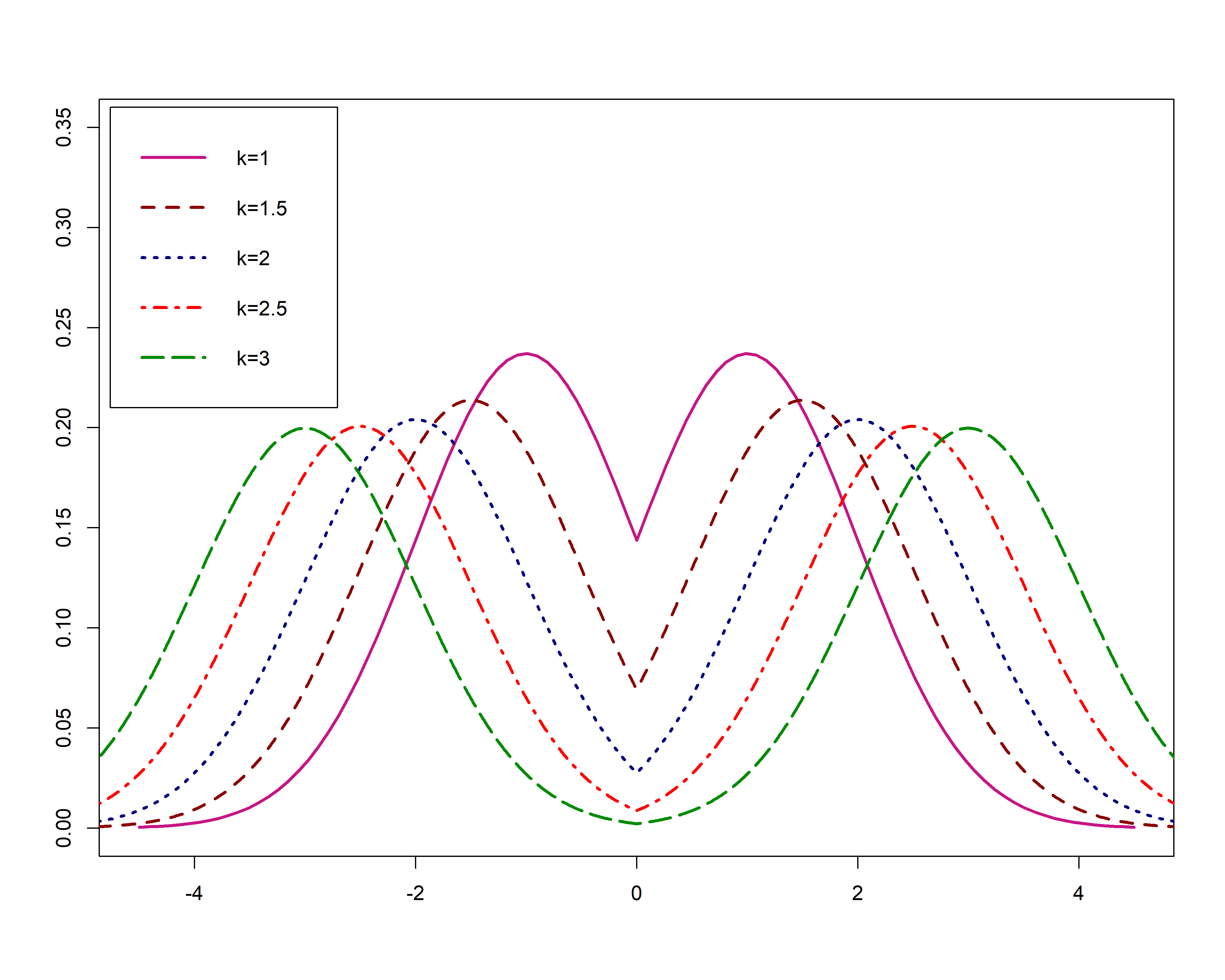}}}\hspace{5pt}
\subfloat[]{
\resizebox*{6cm}{!}{\includegraphics{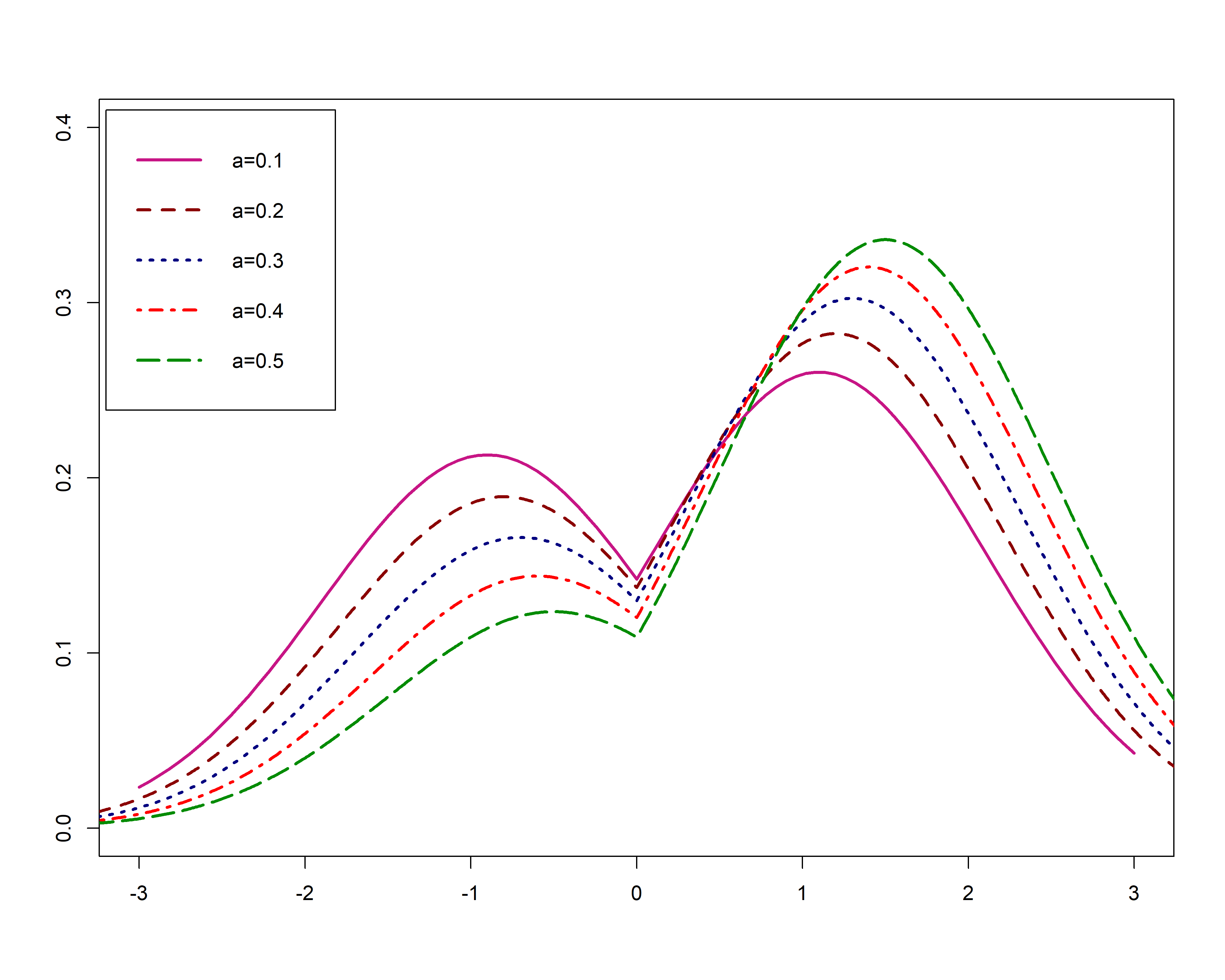}}} \\
\subfloat[]{
\resizebox*{6cm}{!}{\includegraphics{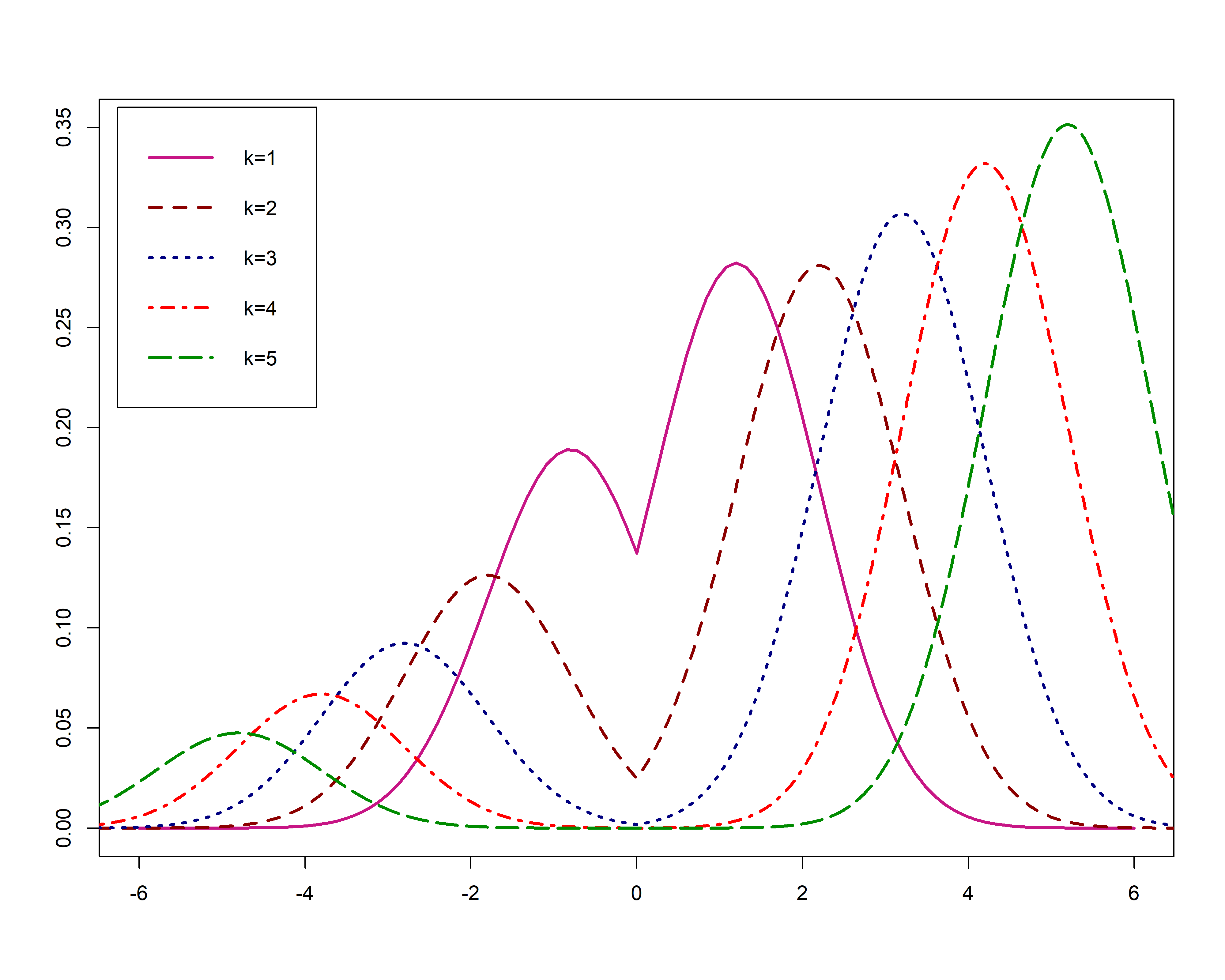}}}\hspace{5pt}
\subfloat[]{
\resizebox*{6cm}{!}{\includegraphics{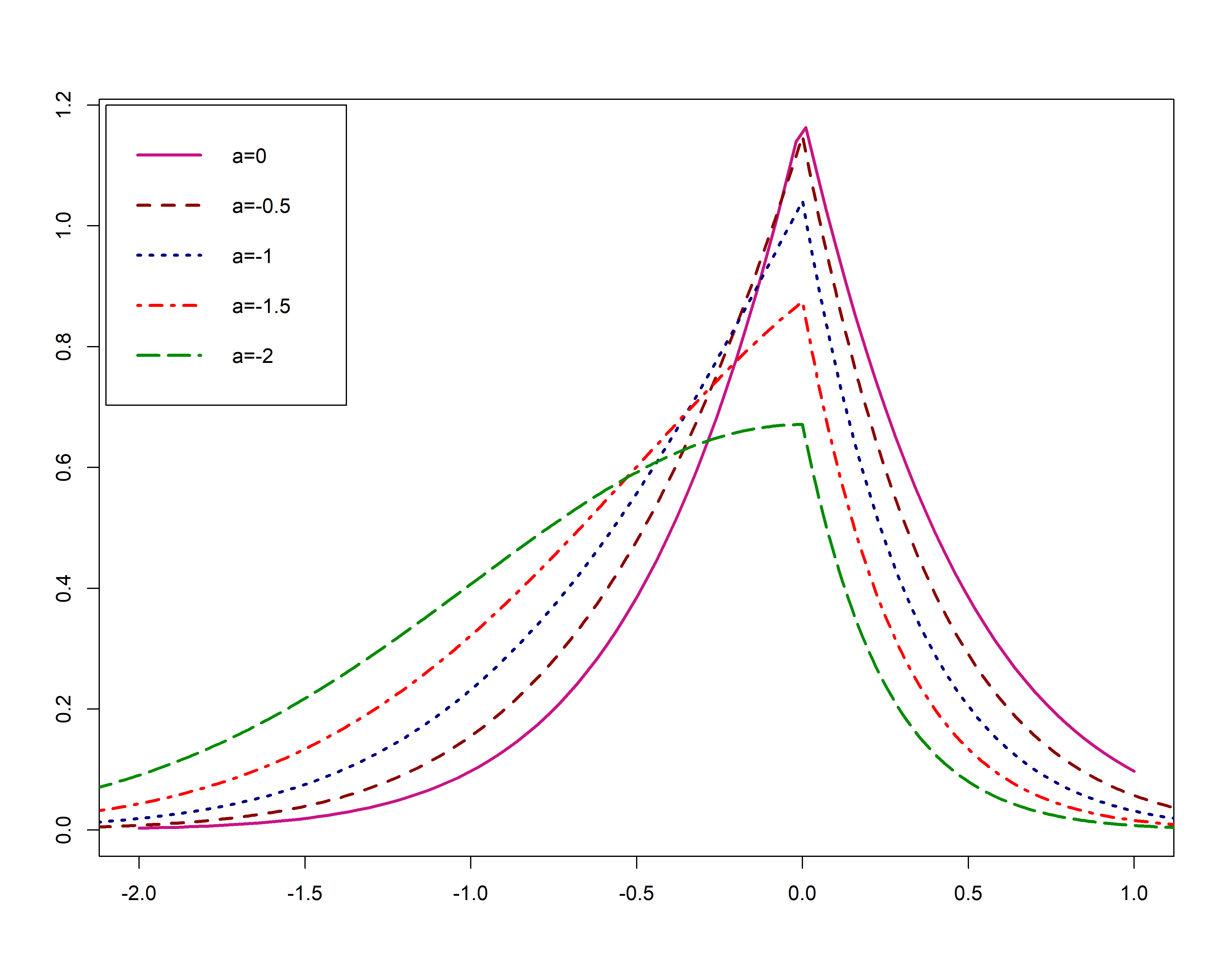}}}
\caption{Shapes of \eqref{eq:11} for some special value of parameters. (a) $ \mu = 0 $, $ \sigma = 1 $ and $ a = 0 $, (b) $ \mu = 0 $, $ \sigma = 1 $ and $ k = 1 $, (c) $ \mu = 0 $, $ \sigma = 1 $ and $ a = 0.2 $ and (d) $ \mu = 0 $, $ \sigma = 1 $ and $ k = -2 $} \label{fig:5}
\end{figure}

In this subsection we will introduce new generalization for normal distribution using theorem \eqref{th:2.4}. the new density function is 
\begin{eqnarray} \label{eq:11}
 f\left(x\right)&=& c_{\sigma, k, a} \exp \left(k\left|\frac{x-\mu }{\sigma }\right|\right) \frac{1}{\sigma} \phi \left(\frac{x-\mu - a }{\sigma }\right)  \qquad x \in \mathbb{R}
\end{eqnarray} 
where $c_{\sigma, k, a}^{-1}=\exp\left( \frac{ka}{\sigma} +\frac{k^2}{2}\right) \mathrm{\Phi }\left(k+\frac{a}{\sigma}\right)+\exp\left( -\frac{ka}{\sigma} +\frac{k^2}{2}\right)\mathrm{\Phi }\left(k-\frac{a}{\sigma}\right)$. If random variable $ X $ has the density function \eqref{eq:11} we say that $X$ has Bimodal-Unimodal Normal (BUN) distribution and denoted by $ X \sim \text{BUN} (\mu, \sigma, k, a) $. $ \mu \in \mathbb{R} $, $ \sigma>0 $ are location and scale, and $ k \in \mathbb{R} $ and $ a \in \mathbb{R} $ are shape parameters. This density function are symmetric for $ a = 0 $ that in this case, we denote by $ X \sim \text{BUN} (\mu, \sigma, k) $. For $ a \neq 0 $ we have skew density function, and for $ k = 0 $ we will have a normal distribution. Bimodal area of \eqref{eq:11} is $ \sigma k > |a| $.
Shapes of density function \eqref{eq:11} for some special value of parameters are given in Figure \eqref{fig:5}.

\begin{proposition} \label{prop1} 
Mode(s) of BUN density function is(are)
\begin{eqnarray}
\left\{ \begin{array}{ll}
          \mu+a\pm\sigma k &  \sigma k > |a| \\ \\
          \mu+a-\sigma k &  a< \sigma k < -a \\ \\
          \mu+a+\sigma k & -a< \sigma k < a \\ \\
          \mu &  \sigma k \leq -|a| 
            .\end{array} \right. \nonumber
\end{eqnarray}       
\end{proposition}

We work on the standard version of BUN, $ Z = \frac{X-\mu}{\sigma} $ that is $ Z \sim \text{BUN} (0, 1, k, \frac{a}{\sigma}) $ for simplicity on calculations. The cumulative distribution function of $ Z $ is given by

\begin{eqnarray}
 F\left(z\right) = \left\{ \begin{array}{ll}
          \frac{{\mathrm{e}}^{-\frac{ka}{\sigma}}\mathrm{\ }\mathrm{\Phi }\left(z+k-\frac{a}{\sigma}\right)}{{\mathrm{e}}^{\frac{ka}{\sigma}}\mathrm{\Phi }\left(k+\frac{a}{\sigma}\right)+{\mathrm{e}}^{-\frac{ka}{\sigma}}\mathrm{\Phi }\left(k-\frac{a}{\sigma}\right)} &  z\leq 0 \\ \\
         \frac{{\mathrm{e}}^{-\frac{ka}{\sigma}}\mathrm{\ }\mathrm{\Phi }\left(k-\frac{a}{\sigma}\right)+{\mathrm{e}}^{\frac{ka}{\sigma}}\left[\mathrm{\Phi }\left(z-k-\frac{a}{\sigma}\right)-\mathrm{\Phi }\left(-k-\frac{a}{\sigma}\right)\right]}{{\mathrm{e}}^{\frac{ka}{\sigma}}\mathrm{\Phi }\left(k+\frac{a}{\sigma}\right)+{\mathrm{e}}^{-\frac{ka}{\sigma}}\mathrm{\Phi }\left(k-\frac{a}{\sigma}\right)} &  z >0 .\end{array} \right. \nonumber
\end{eqnarray}

\subsubsection*{Stochastic representation for BUN}
A key element in BUN construction is that the
distribution can be stochastically represented as a mixture of two truncated random variables without overlap. That means BUN model has a closed-form density function and Furthermore, there is no label switching problem. With applying this fact, we can easily generate sample data from BUN distribution.
\begin{proposition} \label{prop2}
The density function of BUN is a mixture of two truncated normal as
\begin{eqnarray} 
p \frac{\phi \left(z+k-\frac{a}{\sigma}\right)}{\Phi\left(k-\frac{a}{\sigma}\right)} + (1-p) \frac{\phi \left(z-k-\frac{a}{\sigma}\right)}{\Phi\left(k+\frac{a}{\sigma}\right)} \nonumber
\end{eqnarray} 
where $p = \frac{{\text{e}}^{-\frac{ka}{\sigma}}\Phi \left(k-\frac{a}{\sigma}\right)}{{\text{e}}^{\frac{ka}{\sigma}}\Phi\left(k+\frac{a}{\sigma}\right)+{\text{e}}^{-\frac{ka}{\sigma}}\Phi\left(k-\frac{a}{\sigma}\right)}$. It is easy to show that $\frac{\phi \left(z+k-\frac{a}{\sigma}\right)}{\Phi\left(k-\frac{a}{\sigma}\right)}$ and $\frac{\phi \left(z-k-\frac{a}{\sigma}\right)}{\Phi\left(k+\frac{a}{\sigma}\right)}$ are density function of a truncated normal distribution on interval $ (-\infty, 0) $ and $ (0, \infty) $, respectively.    
\end{proposition}   

Moment generating function of $ Z $ is given by
\begin{eqnarray}  
M_Z\left(t\right) = \frac{{\mathrm{e}}^{\left(k+t\right)\frac{a}{\sigma}+\frac{{\left(k+t\right)}^2}{2}}\mathrm{\Phi }\left(k+\frac{a}{\sigma}+t\right)+{\mathrm{e}}^{-\left(k-t\right)\frac{a}{\sigma}+\frac{{\left(k-t\right)}^2}{2}}\mathrm{\Phi }\left(k-\frac{a}{\sigma}-t\right)}{{\mathrm{e}}^{\frac{ka}{\sigma}+\frac{k^2}{2}}\mathrm{\Phi }\left(k+\frac{a}{\sigma}\right)+{\mathrm{e}}^{-\frac{ka}{\sigma}+\frac{k^2}{2}}\mathrm{\Phi }\left(k-\frac{a}{\sigma}\right)} \cdot \nonumber
\end{eqnarray} 
We can easily calculate the moments of the standard BUN distribution by derivative $ M_Z\left(t\right) $ on $ t $. Some moments of $ Z $ are given by
\begin{eqnarray}
E\left(Z\right) &=& \frac{p_k p_t - n_k n_t}{\delta} \nonumber \\
E\left(Z^2\right) &=& \frac{(1+p_k^2)p_t + (1+n_k^2)n_t + 2k n_d }{\delta}  \nonumber \\ 
E\left(Z^3\right) &=& \frac{(3p_k+p_k^3)p_t - (3n_k+n_k^3)n_t + 4\frac{ka}{\sigma} n_d }{\delta} \nonumber \\
E\left(Z^4\right) &=&
\frac{\left(3+6p_k^2+p_k^4\right)p_t + \left(3+6n_k^2+n_k^4\right)n_t + \left[2k^3+10k+6(\frac{a}{\sigma})^2k\right] n_d}{\delta} \nonumber 
\end{eqnarray}
where
\begin{eqnarray}
p_t &=& \text{e}^{\frac{ka}{\sigma}} \Phi\left( k+\frac{a}{\sigma} \right) \nonumber \\
n_t &=& \text{e}^{\frac{-ka}{\sigma}} \Phi\left( k-\frac{a}{\sigma} \right) \nonumber \\
\delta &=& p_t + n_t \nonumber \\
p_k &=&  k+\frac{a}{\sigma}  \nonumber \\  
n_k &=&  k-\frac{a}{\sigma}  \nonumber  \\
n_d &=& \text{e}^{\frac{-ka}{\sigma}} \phi\left( k-\frac{a}{\sigma} \right) \nonumber 
\end{eqnarray}

\subsubsection*{ML estimation of BUN distribution}
Let  $\boldsymbol{\theta}^{BUN} =\left( \mu,\sigma ,k,a\right)^T $  parameters in model and log-likelihood of the model are given by     
\[\ell = \ell \left(\boldsymbol{\theta}^{BUN} \right)=-\frac{nk^2}{2}-n{\log \left(\sigma \right)\ }-n{\log \left(\delta\right)\ }-\frac{n}{2}{\log \left(2\pi \right)\ }+\frac{k}{\sigma }\sum^n_{i=1}{\left|x_i-\mu\right|}-\frac{1}{2}\sum^n_{i=1}{{\left(\frac{x_i-\mu-a}{\sigma }\right)}^2}\] 
To maximize $\ell$ with respect to the parameters of the model, we solve score vector $\boldsymbol{U}^{BUN}_n={\left(\frac{\partial \ell }{\partial \mu },\frac{\partial \ell }{\partial \sigma },\frac{\partial \ell }{\partial k},\frac{\partial \ell }{\partial a}\right)}^T = 0$. Components of score vector are given in the appendix. Solving of this equations does not have closed form, so we proceed through a numerical optimization methods.

\subsection{Bimodal-Unimodal Student-t Distribution}
\begin{figure}  [h!]
\centering
\subfloat[]{
\resizebox*{6cm}{!}{\includegraphics{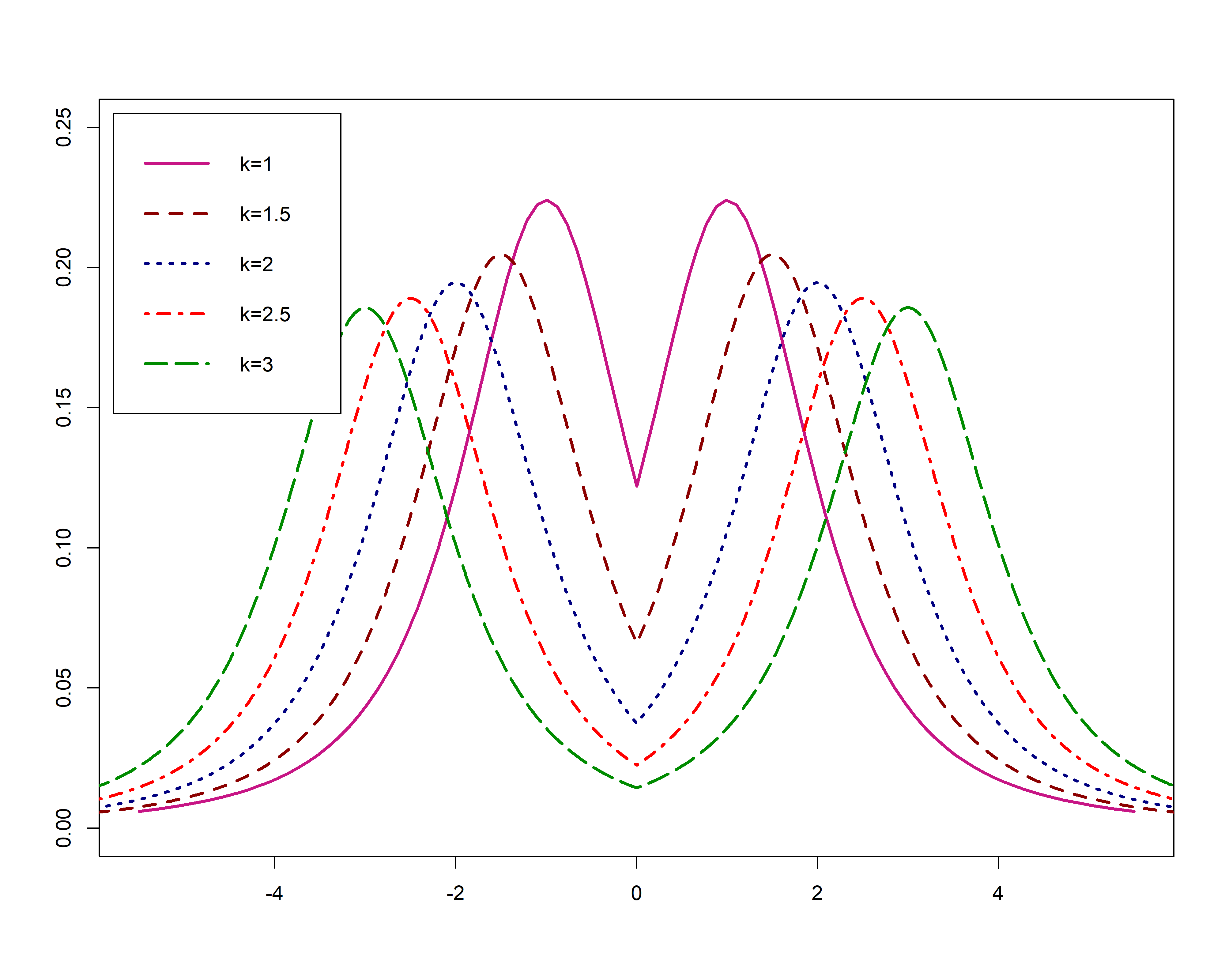}}}\hspace{5pt}
\subfloat[]{
\resizebox*{6cm}{!}{\includegraphics{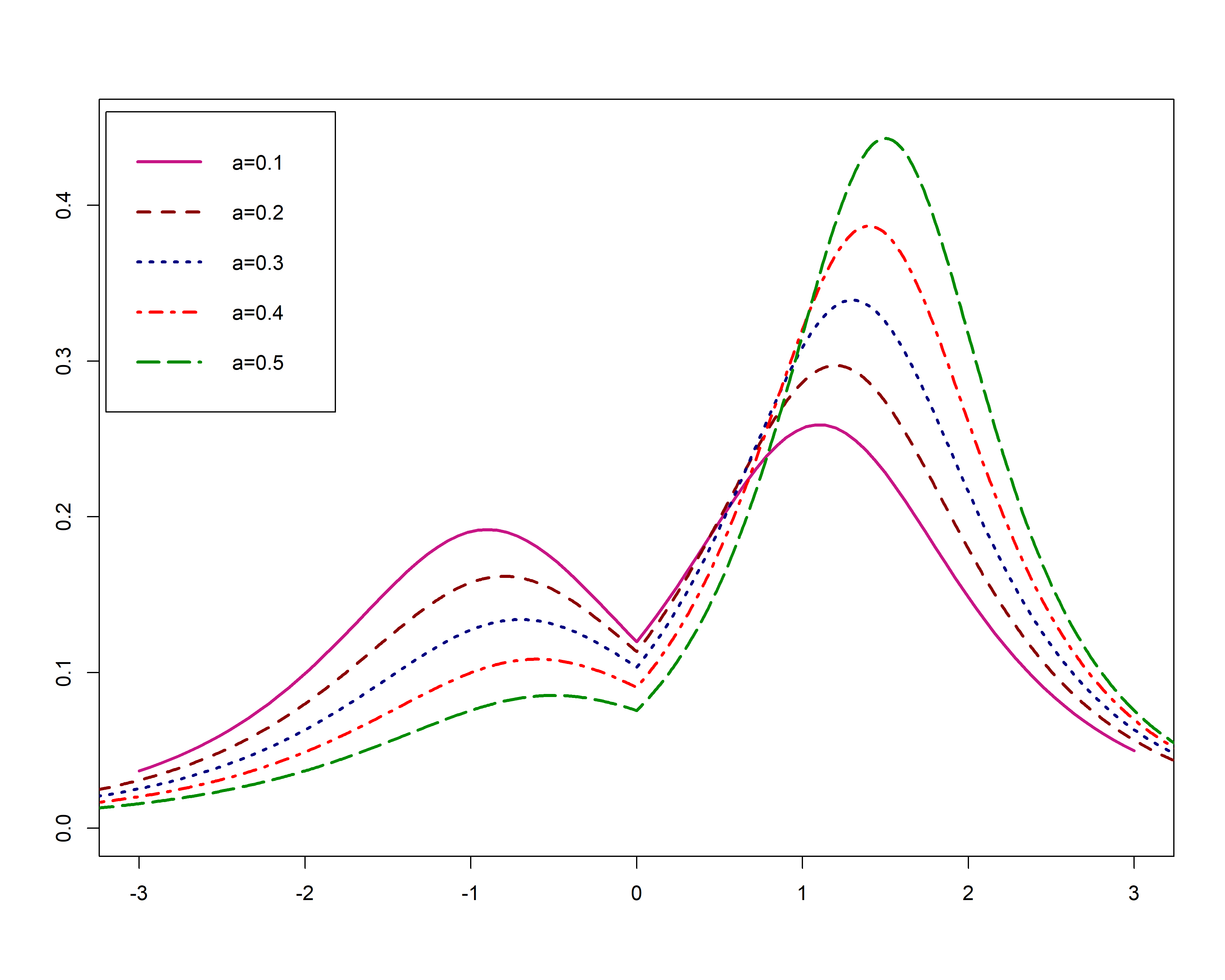}}} \\
\subfloat[]{
\resizebox*{6cm}{!}{\includegraphics{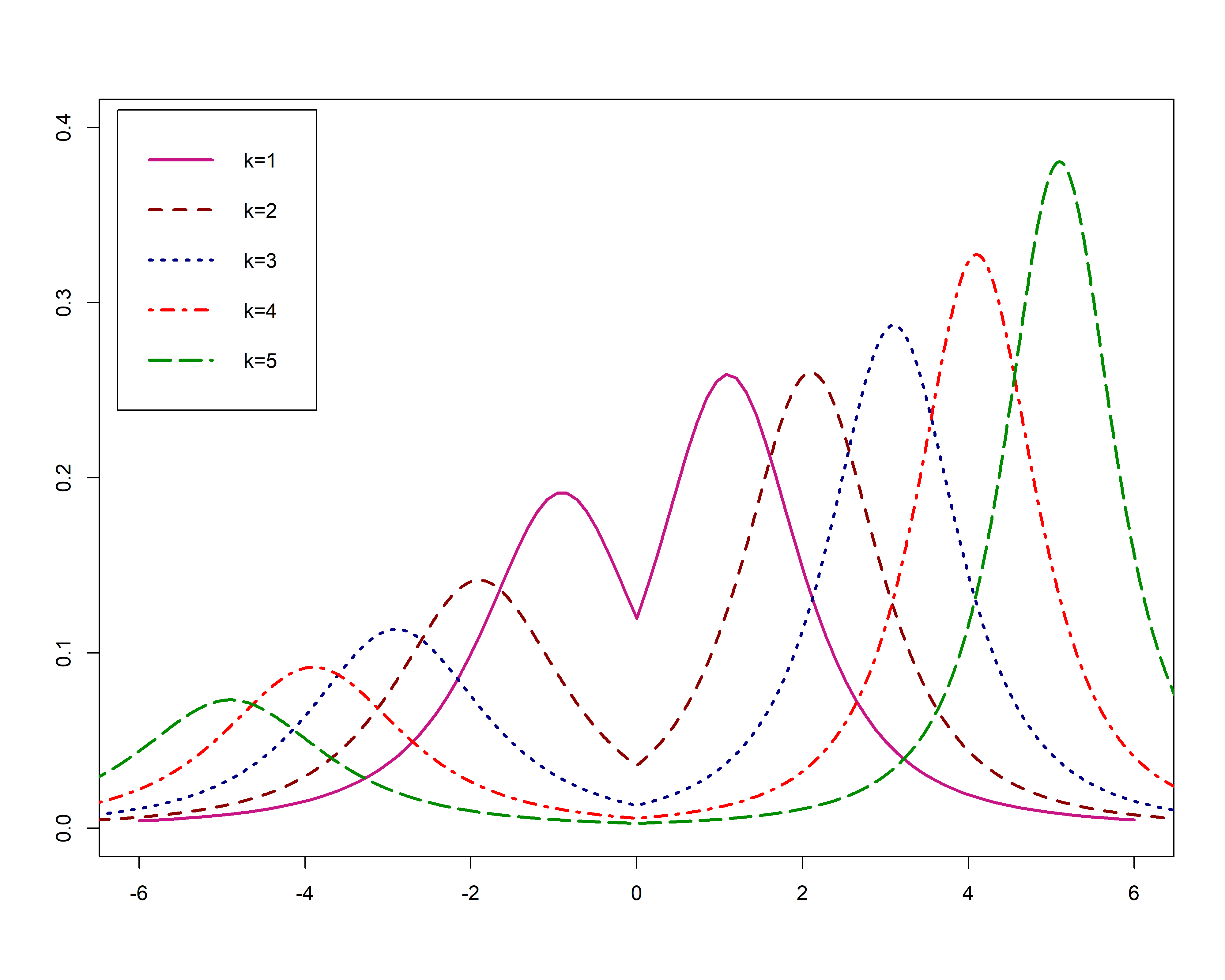}}}\hspace{5pt}
\subfloat[]{
\resizebox*{6cm}{!}{\includegraphics{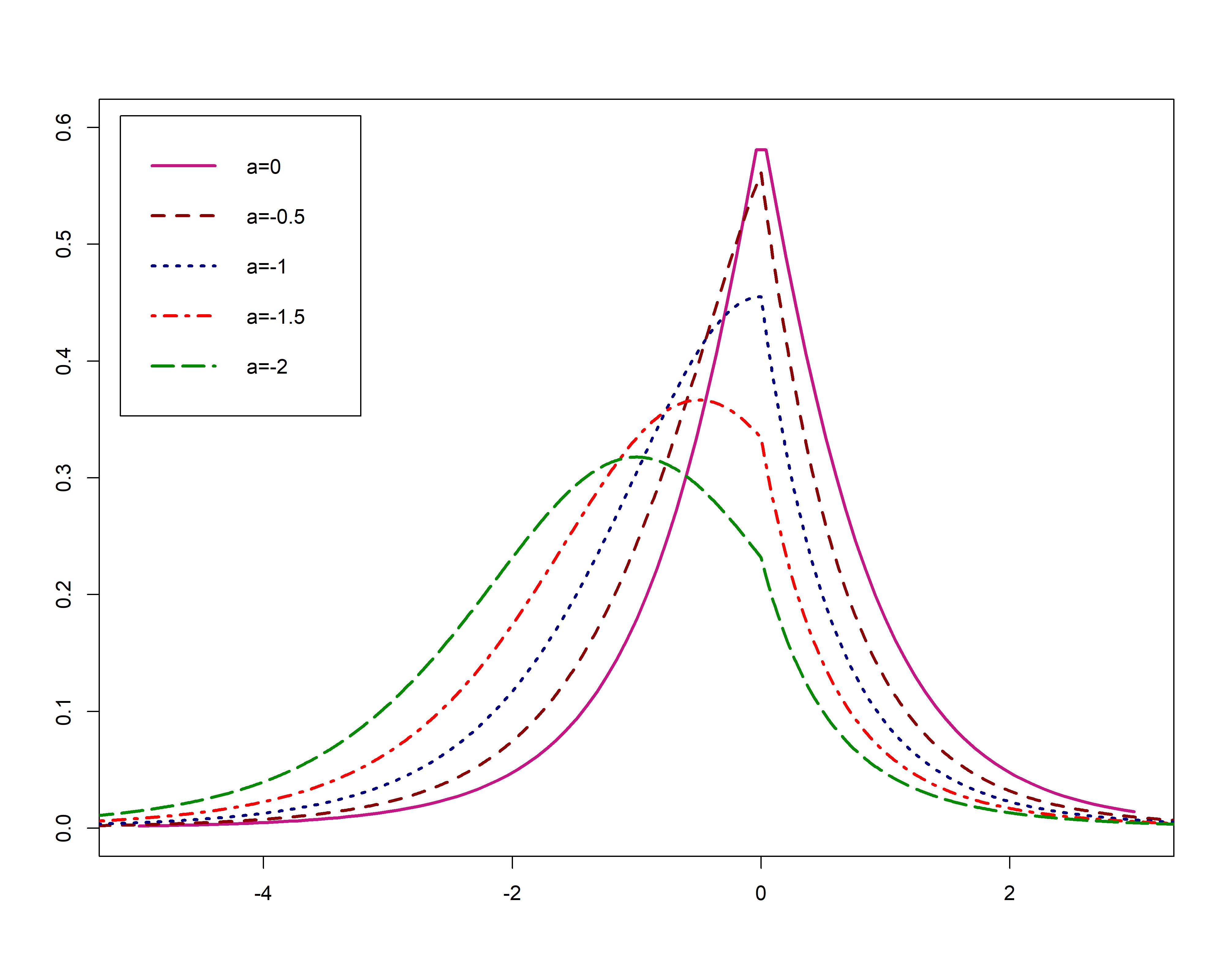}}}
\caption{Shapes of \eqref{eq:13} for some special value of parameters. (a) $ \mu = 0 $, $ \sigma = 1 $, $ a = 0 $ and $ \nu = 2 $, (b) $ \mu = 0 $, $ \sigma = 1 $, $ k = 1 $ and $ \nu = 2 $, (c) $ \mu = 0 $, $ \sigma = 1 $, $ a = 0.1 $ and $ \nu = 2 $ and (d) $ \mu = 0 $, $ \sigma = 1 $, $ k = -1 $ and $ \nu = 5 $} \label{fig:6}
\end{figure}

The most popular distribution which is suitable for heavy-tail data is student-t. There are some extension of this distribution. For example, see Jones and Faddy (2003), Aas and Haff (2006), Nadarajah and Kotz (2006), Balakrishnan (2009) and Huang et al. (2019). In this section, we will work on new generalization for student-t distribution using theorem \eqref{th:2.5}. The new density function is 
\begin{eqnarray} \label{eq:12}
f(x) = \frac{d_{\nu}\left(\left|\frac{x-\mu}{\sigma}\right| - k \right) } {2 \sigma D_{\nu}\left( k\right)  } =  \frac{ \left\{d_{\nu}\left(\frac{x-\mu}{\sigma} - k \right) \right\} ^ {I\left( x \geq \mu\right)}  \left\{d_{\nu}\left(\frac{x + \mu}{\sigma} + k \right) \right\} ^ {I\left( x < \mu\right)} } {2 \sigma D_{\nu}\left( k\right)  }
\end{eqnarray}
that is a Bimodal density function for $ k >0 $, unimodal for $ k <0 $ and student-t distribution for $ k = 0 $. This distribution is symmetric and is not usable for skew and asymmetric data. We can extend this distribution to 
\begin{eqnarray} \label{eq:13}
f(x) = \frac{\left\{s_{-} ^ {-\frac{\nu + 1}{2}} d_{\nu}\left( \frac{x - \mu - \left( a + k\right) }{\sqrt{s_{-}}}\right) \right\} ^ {I\left( x \geq \mu\right) }   \left\{s_{+} ^ {-\frac{\nu + 1}{2}} d_{\nu}\left( \frac{x - \mu - \left( a - k\right) }{\sqrt{s_{+}}}\right) \right\} ^ {I\left( x < \mu\right)}} {s_{-} ^ {-\frac{\nu}{2}}   D_{\nu}\left( \frac{   a + k }{\sqrt{s_{-}}}\right) + s_{+} ^ {-\frac{\nu}{2}}   D_{\nu}\left( \frac{  k - a }{\sqrt{s_{+}}}\right) }  
\end{eqnarray}
where $ d_{\nu}\left( \cdot \right)  $ and $ D_{\nu} \left( \cdot \right)  $ pdf and cdf of student-t distribution with $\nu$ degrees of freedom and $ s _{-} = \frac{\nu \sigma^2 - 2ak}{\nu}  $ and $ s_{+} = \frac{\nu \sigma^2 + 2ak}{\nu} $. If random variable $X$ has density function \eqref{eq:13} we say $X$ follow Bimodal-Unimodal student t (BUSt) distribution and write $X \sim BUSt(\mu, \sigma, k, a, \nu)$. When $a=0$ the density function \eqref{eq:13} is equivalent to the \eqref{eq:12} and in this case we write $X \sim BUSt(\mu, \sigma, k, \nu)$. Shapes of density function $ \eqref{eq:13} $ are shown in  Figure \eqref{fig:6}. 

\begin{proposition} \label{prop3}
Mode(s) of BUSt distribution are
\begin{eqnarray}
\left\{ \begin{array}{ll}
          \mu+a\pm k &   k > |a| \\ \\
          \mu+a- k &  a<  k < -a \\ \\
          \mu+a+ k & -a<  k < a \\ \\
          \mu &  k \leq -|a| 
            .\end{array} \right. \nonumber
\end{eqnarray}             
\end{proposition}
\begin{proposition} \label{prop4}
If $X \sim BUSt(\mu, \sigma, k, a, \nu)$ , then for $\nu \rightarrow \infty$
\begin{equation}
X \overset{d}{\longrightarrow} Y \nonumber
\end{equation}
where $ Y \sim BUN(\mu, \sigma, k, a)$ and $ \overset{d}{\longrightarrow} $ means convergence in distribution.       
\end{proposition}

\subsubsection*{Stochastic representation for BUSt}
Like the BUN, the BUSt distribution has stochastic representation as a mixture of two truncated random variables without overlap. So with using this fact, we can generate sample data from BUSt distribution and calculate moments of it.
\begin{proposition} \label{prop5}
Density \eqref{eq:13} is a mixture of two truncated Student t distributions, that is
 \begin{eqnarray}
 f\left(x\right) = R\left(s_{-} \right) f\left( x_1\right)  + R\left( s_{+}\right) f\left( x_2\right) \nonumber
 \end{eqnarray}         
Where $ X_1 \sim Tt_{(\mu, \infty)} \left( \mu + (a+k) , \sqrt{s_-} ; \nu\right)  $, $ X_1 \sim Tt_{(-\infty, \mu)} \left( \mu + (a-k) , \sqrt{s_+} ; \nu\right) $ and
\begin{eqnarray}
R\left(s_{-} \right) &=& \frac{ s_{-} ^ {-\frac{\nu}{2}}   D_{\nu}\left( \frac{   a + k }{\sqrt{s_{-}}}\right) } {s_{-} ^ {-\frac{\nu}{2}}   D_{\nu}\left( \frac{   a + k }{\sqrt{s_{-}}}\right) + s_{+} ^ {-\frac{\nu}{2}}   D_{\nu}\left( \frac{  k - a }{\sqrt{s_{+}}}\right) } \nonumber \\
R\left( s_{+}\right) &=& \frac{ s_{+} ^ {-\frac{\nu}{2}}   D_{\nu}\left( \frac{   k - a }{\sqrt{s_{+}}}\right) } {s_{-} ^ {-\frac{\nu}{2}}   D_{\nu}\left( \frac{   a + k }{\sqrt{s_{-}}}\right) + s_{+} ^ {-\frac{\nu}{2}}   D_{\nu}\left( \frac{  k - a }{\sqrt{s_{+}}}\right) } \nonumber
\end{eqnarray}
From Kim (2008), density function $ X_1 $ and $ X_2 $ are given by
\begin{eqnarray}
f\left( x_1\right)  &=& \frac{s_{-} ^ {-\frac{1}{2}} d_{\nu}\left( \frac{x_1 - \mu - \left( a + k\right) }{\sqrt{s_{-}}}\right)}{D_{\nu}\left( \frac{   a + k }{\sqrt{s_{-}}}\right)} \nonumber \\
f\left( x_2\right) &=& \frac{s_{+} ^ {-\frac{1}{2}} d_{\nu}\left( \frac{x_2 - \mu - \left( a - k\right) }{\sqrt{s_{+}}}\right)}{D_{\nu}\left( \frac{   k - a }{\sqrt{s_{+}}}\right)} \nonumber 
\end{eqnarray}
\end{proposition}

We can use proposition \eqref{prop5} to generating random sample and calculate moments of BUSt distribution. 
The $r$-th moment of $ X \sim BUSt(\mu, \sigma, k, a, \nu) $ for $ r = 1,2,3,4 $ is given by
\begin{eqnarray}
E\left( X^r\right)  &=& \sum_{i = 0}^{r} \binom{r}{i} R(s_-)^i R(s_+)^{r-i} E\left( X_1 ^ i\right) E\left( X_2 ^ {r-i}\right)  \nonumber 
\end{eqnarray}
According to Kim (2008), we have that
\begin{eqnarray}
E\left( X_1^m\right)  = \sum_{j = 0}^{m} \binom{m}{j} \left( \mu + a + k\right) ^ {m - j} {\left( s_-\right) } ^ {\frac{j}{2}} \eta_{m}  \nonumber 
\end{eqnarray}
And 
\begin{eqnarray}
E\left( X_2^l\right)  = \sum_{j = 0}^{l} \binom{l}{j} \left( \mu + a - k\right) ^ {l - j} {\left( s_+\right) } ^ {\frac{j}{2}} \lambda_{l}  \nonumber 
\end{eqnarray}
For $ m, l = 1, 2, 3, 4 $, where 
\begin{eqnarray}
\eta_{1} &=& G_{\nu}(1) \left( \nu + \frac{\left( a + k \right)  ^ 2}{{s_-}} \right) ^ {-\frac{\nu - 1}{2}}  \quad \nu > 1  \nonumber \\
\eta_{2} &=& \frac{\nu}{\nu - 2} -  \frac{a + k}{\sqrt{s_-}} G_{\nu}(1) \left( \nu + \frac{\left( a + k \right)  ^ 2}{{s_-}} \right) ^ {-\frac{\nu - 1}{2}} \quad \nu > 2  \nonumber \\
\eta_{3} &=& G_{\nu}(3) \left( \nu + \frac{\left( a + k \right)  ^ 2}{{s_-}} \right) ^ {-\frac{\nu - 3}{2}} + \frac{\left( a + k\right) ^2}{{s_-}} G_{\nu}(1) \left( \nu + \frac{\left( a + k \right)  ^ 2}{{s_-}} \right) ^ {-\frac{\nu - 1}{2}} \quad \nu > 3  \nonumber \\
\eta_{4} &=& 3 \left\{ \frac{\nu^2}{(\nu-2)(\nu-4)} - \frac{G_{\nu}(3)}{2} \frac{a + k}{\sqrt{s_-}} \left( \nu + \frac{\left( a + k \right)  ^ 2}{{s_-}} \right) ^ {-\frac{\nu - 3}{2}} \right\}  \nonumber\\
&& - \frac{(a + k)^3}{{\left( s_-\right) }^{\frac{3}{2}}} G_{\nu}(1) \left( \nu + \frac{\left( a + k \right)  ^ 2}{{s_-}} \right) ^ {-\frac{\nu - 1}{2}} \quad \nu > 4  \nonumber
\end{eqnarray}
and
\begin{eqnarray}
\lambda_{1} &=& -G'_{\nu}(1) \left( \nu + \frac{\left( a - k \right)  ^ 2}{{s_+}} \right) ^ {-\frac{\nu - 1}{2}}  \quad \nu > 1  \nonumber \\
\lambda_{2} &=& \frac{\nu}{\nu - 2} +  \frac{a - k}{\sqrt{s_+}} G'_{\nu}(1) \left( \nu + \frac{\left( a - k \right)  ^ 2}{{s_+}} \right) ^ {-\frac{\nu - 1}{2}} \quad \nu > 2  \nonumber \\
\lambda_{3} &=& - G'_{\nu}(3) \left( \nu + \frac{\left( a - k \right)  ^ 2}{{s_+}} \right) ^ {-\frac{\nu - 3}{2}} - \frac{\left( a - k\right) ^2}{{s_+}} G'_{\nu}(1) \left( \nu + \frac{\left( a - k \right)  ^ 2}{{s_+}} \right) ^ {-\frac{\nu - 1}{2}} \quad \nu > 3  \nonumber \\
\lambda_{4} &=& 3 \left\{ \frac{\nu^2}{(\nu-2)(\nu-4)} + \frac{G'_{\nu}(3)}{2} \frac{a - k}{\sqrt{s_+}} \left( \nu + \frac{\left( a - k \right)  ^ 2}{{s_+}} \right) ^ {-\frac{\nu - 3}{2}} \right\}  \nonumber\\
&& + \frac{(a - k)^3}{{\left( s_+\right) }^{\frac{3}{2}}} G'_{\nu}(1) \left( \nu + \frac{\left( a - k \right)  ^ 2}{{s_+}} \right) ^ {-\frac{\nu - 1}{2}} \quad \nu > 4  \nonumber
\end{eqnarray}
and for $ s = 1, 2 $
\begin{eqnarray}
G_{\nu}(s) = \frac{\Gamma\left( \frac{\nu - s}{2} \right) \nu ^ {\frac{\nu}{2}} }{2 D_{\nu} \left( \frac{a + k}{\sqrt{s_-}}\right) \Gamma\left( \frac{\nu}{2} \right)  \Gamma\left( \frac{1}{2} \right)  }  \nonumber
\end{eqnarray}
and
\begin{eqnarray}
G'_{\nu}(s) = \frac{\Gamma\left( \frac{\nu - s}{2} \right) \nu ^ {\frac{\nu}{2}} }{2 D_{\nu} \left( \frac{k - a}{\sqrt{s_+}}\right) \Gamma\left( \frac{\nu}{2} \right)  \Gamma\left( \frac{1}{2} \right)  }  \nonumber
\end{eqnarray}
Cumulative distribution function of $ \eqref{eq:15} $ is given by

\begin{eqnarray} \label{eq:14}
F\left(x\right) = \left\{ \begin{array}{ll}
          \frac{  s_{+} ^ {-\frac{\nu }{2}} D_{\nu}\left( \frac{x - \mu - \left( a - k\right) }{\sqrt{s_{+}}}\right)}{s_{-} ^ {-\frac{\nu}{2}}   D_{\nu}\left( \frac{   a + k }{\sqrt{s_{-}}}\right) + s_{+} ^ {-\frac{\nu}{2}}   D_{\nu}\left( \frac{  k - a }{\sqrt{s_{+}}}\right) }  &  x \leq \mu \\ \\
         \frac{s_{-} ^ {-\frac{\nu }{2}} \left( D_{\nu}\left( \frac{x - \mu - \left( a + k\right) }{\sqrt{s_{-}}}\right) - D_{\nu}\left( \frac{ -k - a }{\sqrt{s_{-}}}\right)\right)   +  s_{+} ^ {-\frac{\nu }{2}} D_{\nu}\left( \frac{  k - a }{\sqrt{s_{+}}}\right) }{s_{-} ^ {-\frac{\nu}{2}}   D_{\nu}\left( \frac{   a + k }{\sqrt{s_{-}}}\right) + s_{+} ^ {-\frac{\nu}{2}}   D_{\nu}\left( \frac{  k - a }{\sqrt{s_{+}}}\right) } &  x > \mu \end{array} \right.
\end{eqnarray}         

\subsubsection*{ML estimation of BUSt distribution}
Let  $\boldsymbol{\theta}^{BUSt} =(\mu ,\sigma ,k, a, \nu)^T$  parameters in the BUSt model and log-likelihood of the model are given by     
\begin{eqnarray}
\ell=\ell \left(\boldsymbol{\theta}^{BUSt} \right) &=& -n\log\left( \delta \right) - \dfrac{\nu + 1}{2}  \log\left(s_{-} \right) \sum_{i = 1}^{n} I\left(x_i \geq \mu \right) - \dfrac{\nu + 1}{2} \log\left(s_{+} \right) \sum_{i = 1}^{n} I\left(x_i < \mu \right)  \nonumber \\
&&  + n \log\left(\Gamma\left( \dfrac{\nu + 1}{2}\right)  \right) - n \log\left(\Gamma\left( \dfrac{\nu }{2}\right)  \right) - n \log\left(\sqrt{\nu \pi} \right)  \nonumber  \\ 
&&- \dfrac{\nu + 1}{2} \sum_{i = 1}^{n} \left\{ \log\left( 1 + \dfrac{1}{\nu} \left( \dfrac{u^{-}_i}{\sqrt{s_{-}}}\right) ^ 2 \right) I\left(x_i \geq \mu \right)  \right\}   \nonumber \\
&& - \dfrac{\nu + 1}{2} \sum_{i = 1}^{n} \left\{ \log\left( 1 + \dfrac{1}{\nu} \left( \dfrac{u^{+}_i}{\sqrt{s_{+}}}\right) ^ 2 \right)  I\left(x_i < \mu \right) \right\} \nonumber
\end{eqnarray}
Component of the score vector
$\boldsymbol{U}^{BUSt}_n=\left( \frac{\partial \ell }{\partial \mu}, \frac{\partial \ell }{\partial \sigma}, \frac{\partial \ell }{\partial k}, \frac{\partial \ell }{\partial a}, \frac{\partial \ell }{\partial \nu}\right)^T $ 
are given in the appendix. Like BUN model, Solving of this equations does not have closed form, so we proceed through a numerical optimization methods.
\subsection{Bimodal-Unimodal Laplace (BUL) Distribution}
Laplace distribution is famous semi-heavy-tail density function and more applicable one. There are many generalizations of this distribution in the literature. For example see Koenker and Machado (1999), Aryal and Nadarajah (2005), Nekoukhou and Alamatsaz (2012), Shams and Alamatsaz (2013), Yilmaz (2014) and Shah et al. (2019). In this subsection, we extend the Laplace distribution to the more flexible one that is usable in bimodal, unimodal, symmetric and skewed datasets. The density function of new distribution is given by
\begin{eqnarray} \label{eq:15}
f_{BUL}(x) = \frac{k}{\sigma c } \left( 1 + \left( u-\frac{a}{\sigma}\right) ^ 2 \right)  \text{e}^{-k\left| u \right|} 
\end{eqnarray}
If random variable $X$ has density function \eqref{eq:15} we say $X$ follow Bimodal-Unimodal Laplace (BUL) distribution and denoted by $ X \sim \text{BUL}(\mu, \sigma, k, a) $. The BUL density function is symmetric for $ a = 0 $ that in this case, we denote by $ X \sim \text{BUL} (\mu, \sigma, k) $. The cumulative distribution function of BUL is given by 
\begin{eqnarray} \label{eq:16}
F_{BUL}\left(x\right) = \left\{ \begin{array}{ll}
          \frac{k}{c} \left(\frac{u^2}{k} -\frac{2}{k} \left( \frac{a}{\sigma} + \frac{1}{k} \right) \left( u-\frac{1}{k} \right) + \left( \frac{a ^2}{\sigma^2k} + \frac{1}{k} \right)  \right) \text{e}^{ku}    &  x \geq \mu \\ \\
         \frac{k}{c} \left(\frac{2}{k^2} \left( \frac{a}{\sigma} + \frac{1}{k} \right) + \left( \frac{a^2}{\sigma^2k} + \frac{1}{k} \right)  \right) \\
          + \frac{k}{c} \left( \left( \frac{a^2}{\sigma^2k} + \frac{1}{k} \right) \left( 1-\text{e}^{-ku} \right)  - \frac{u^2}{k} \text{e}^{-ku}\right) &  x < \mu \\
           - \frac{2k}{c} \left( \frac{a}{\sigma} - \frac{1}{k} \right)  \left( \frac{1-\text{e}^{-ku}-ku\text{e}^{-ku}}{k^2} \right)        .\end{array} \right.
\end{eqnarray}

\begin{figure}  [h!]
\centering
 \subfloat[]{
 \resizebox*{6cm}{!}{\includegraphics{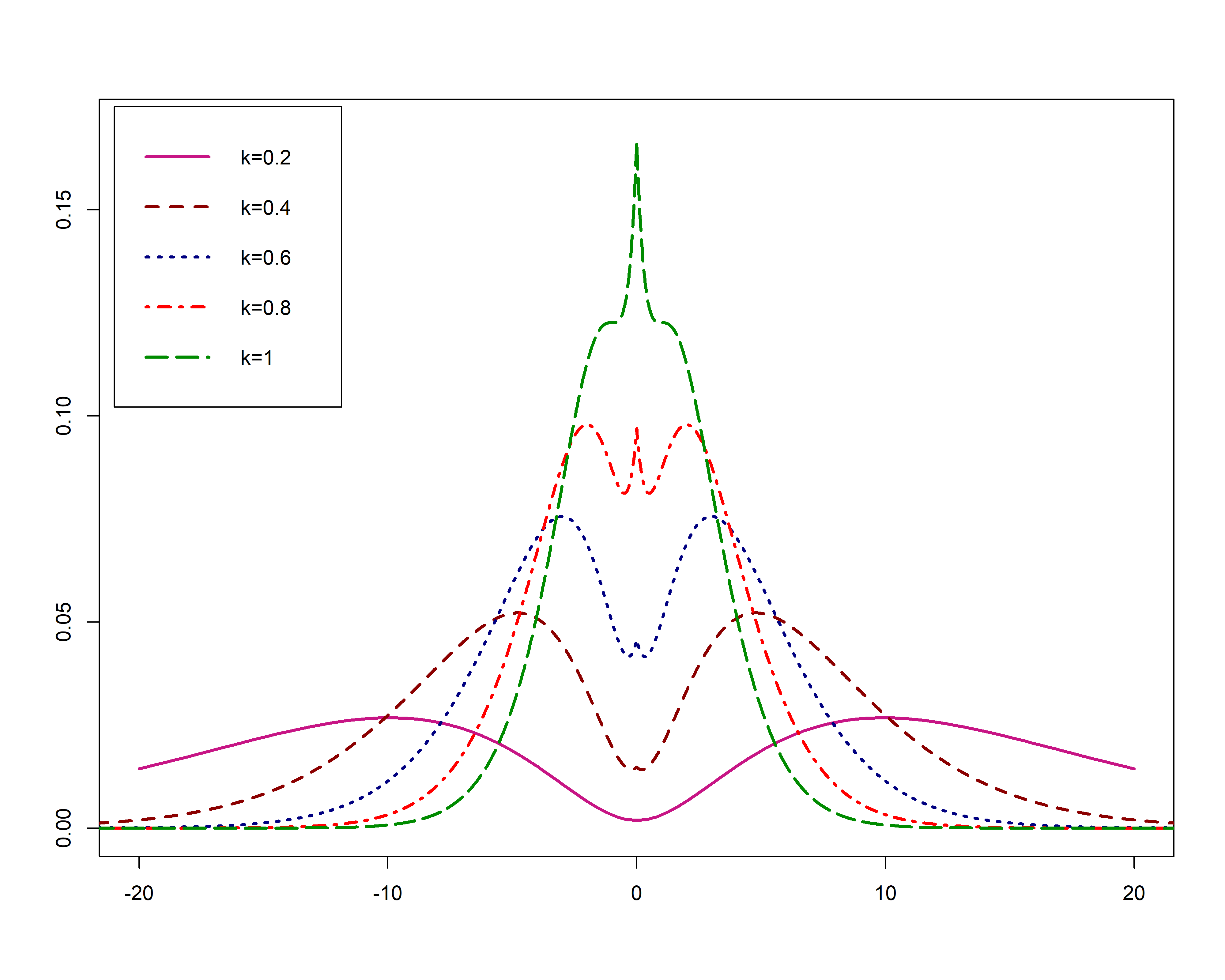}}}\hspace{5pt}
 \subfloat[]{
 \resizebox*{6cm}{!}{\includegraphics{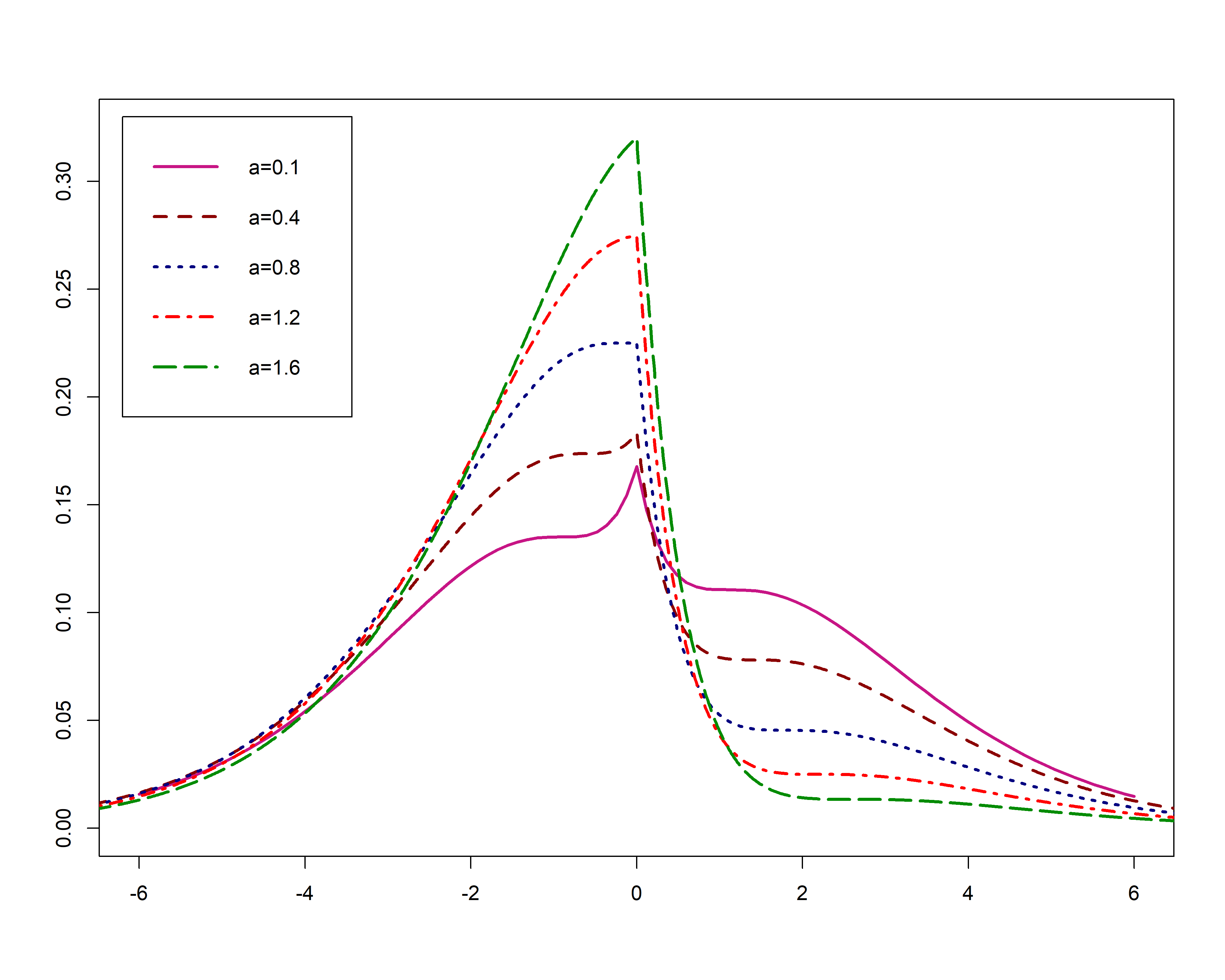}}} \\
 \subfloat[]{
 \resizebox*{6cm}{!}{\includegraphics{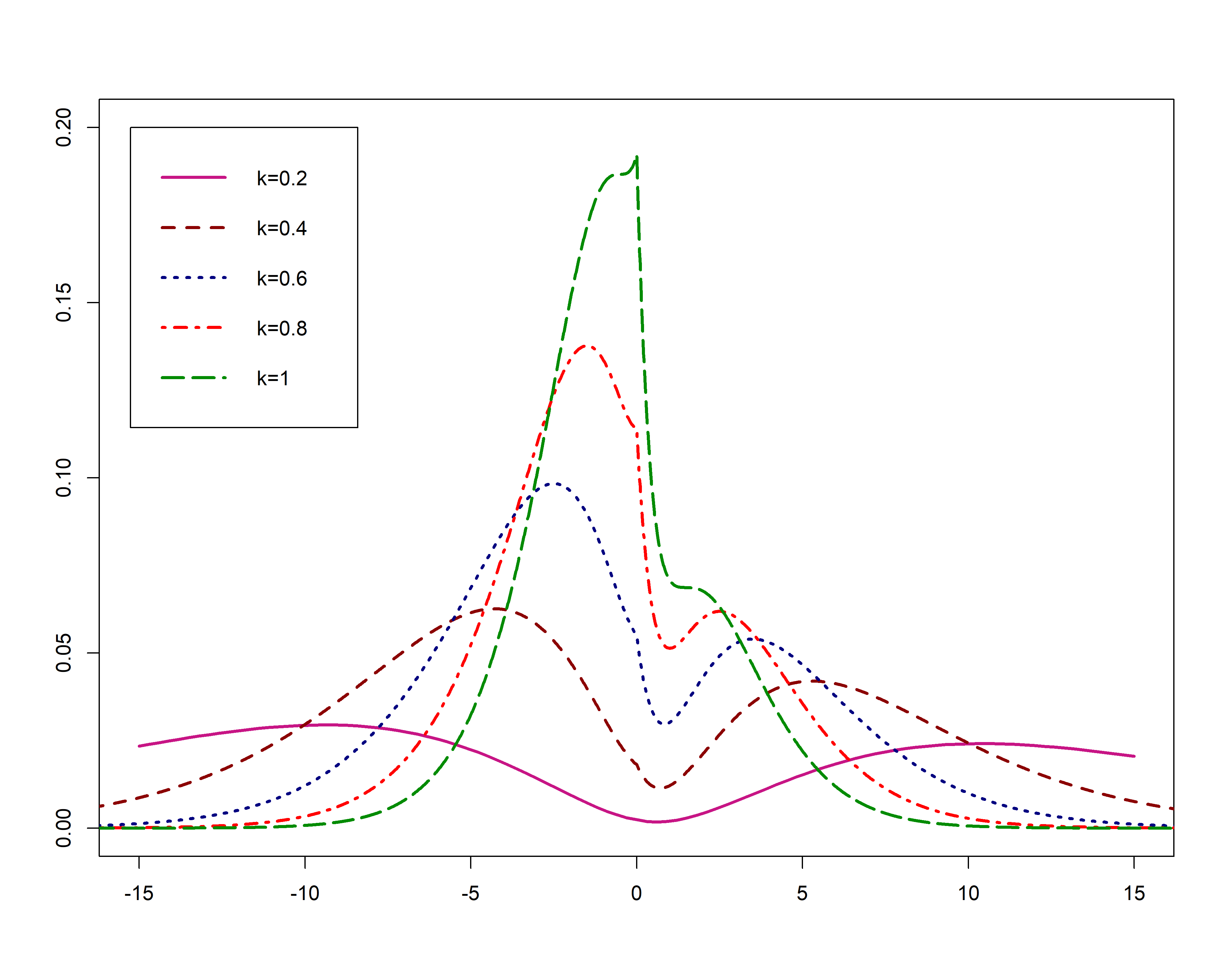}}}\hspace{5pt}
 \subfloat[]{
 \resizebox*{6cm}{!}{\includegraphics{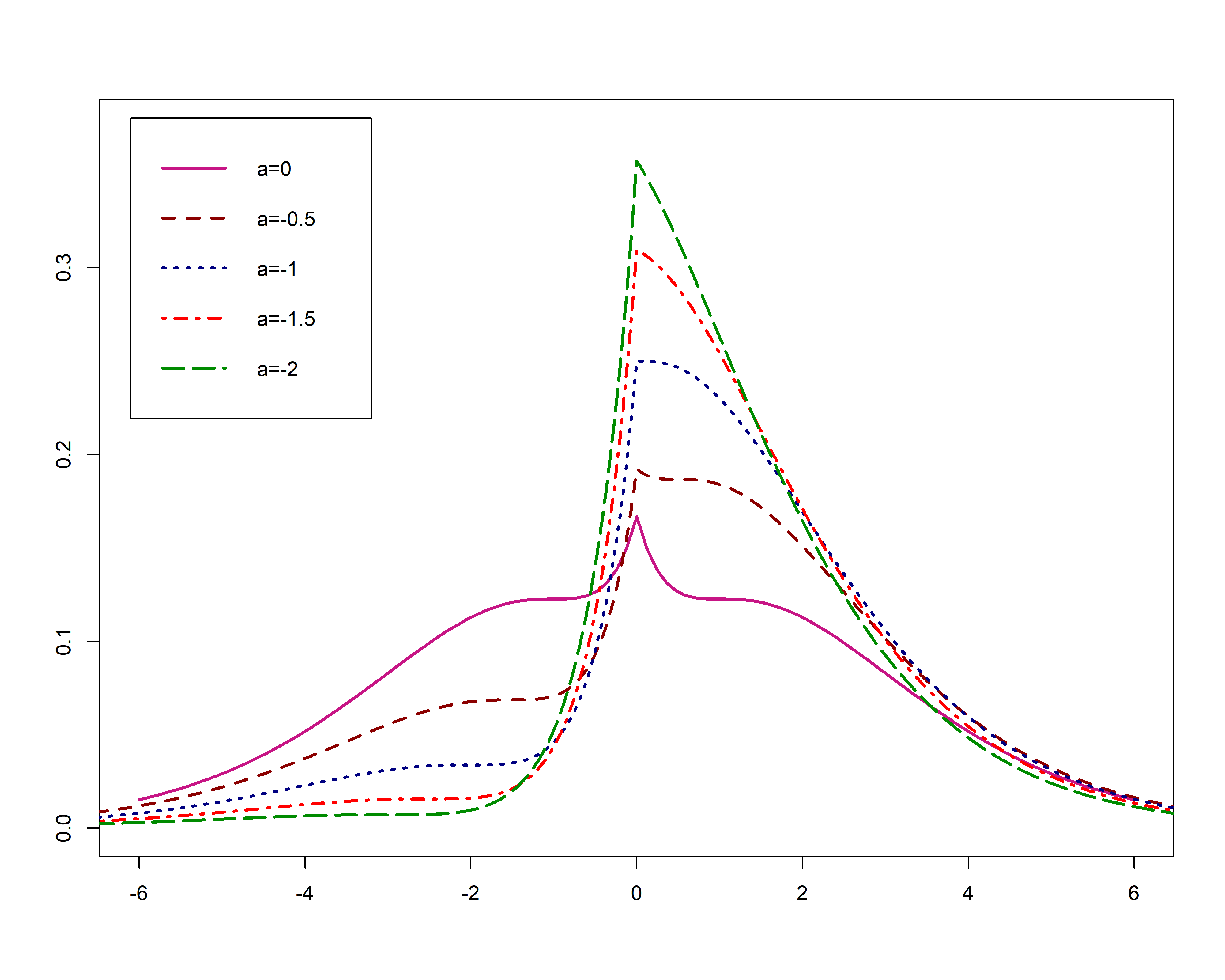}}}
 \caption{Shapes of \eqref{eq:15} for some special value of parameters. (a) $ \mu = 0 $, $ \sigma = 1 $ and $ a = 0 $, (b) $\mu = 0 $, $ \sigma = 1 $ and $ k = 1 $, (c) $ \mu = 0 $, $ \sigma = 1 $ and $ a = 0.5 $ and (d) $ \mu = 0 $, $ \sigma = 1 $ and $ k = 1 $ } \label{fig:7}
 \end{figure}

Where $ c = 2\left( 1+\frac{a^2}{\sigma^2}+\frac{2}{k^2} \right) $ and $ u = \frac{x-\mu}{\sigma}$. Shapes of $ \eqref{eq:15} $ are shown in Figure \eqref{fig:7}.
        
\begin{proposition} \label{prop6}
Modes of BUL are $ \mu+ a \pm \sigma \left( \frac{1}{k} + \sqrt{\frac{1}{k^2}-1} \right)  $ and $\mu$ for $k\leq1$ and $\mu$ otherwise.       
\end{proposition}
Let $ Z=\dfrac{X-\mu}{\sigma} $ standard version of BUL distribution. Then we have that $ Z \sim \text{BUL}(0, 1, k, \frac{a}{\sigma}) $ and the $r$th moment of $Z$ are given by
\begin{equation}
E\left( Z^r \right) = \frac{2}{c} \left\{(1+\frac{a^2}{\sigma^2}) E\left( W_{L} ^ {r}  \right) - 2 \frac{a}{\sigma} E\left( W_{L} ^ {r+1}  \right) + E\left( W_{L} ^ {r+2}  \right)  \right\}  \nonumber
\end{equation}
Where $ W_L $ is Laplace random variable with location zero and scale $\frac{1}{k}$. It is clear that $s$th moment of $ W_L $ is $\dfrac{s!}{2k^s} \left(1+(-1)^s \right) $.
\subsubsection*{ML estimation of BUL distribution}
Parameters in BUL distribution are $\boldsymbol{\theta}^{BUL} =\left(\mu,\sigma ,k,a\right) $ and log-likelihood of BUL are given by     
\[\ell = \ell \left(\boldsymbol{\theta}^{BUL} \right)=n\log\left( \dfrac{k}{2\sigma}\right) -n\log\left( 1+\frac{a^2}{\sigma^2}+\frac{2}{k^2} \right)-k\sum_{i=1}^{n} \left| \dfrac{x_i - \mu}{\sigma}\right| + \sum_{i=1}^{n} \log\left( 1+\left(\dfrac{x_i-\mu-a}{\sigma} \right) ^2 \right) \] 
We prepared components of score vector $\boldsymbol{U}_n^{BUL}={\left(\frac{\partial \ell }{\partial \mu },\frac{\partial \ell }{\partial \sigma },\frac{\partial \ell }{\partial k},\frac{\partial \ell }{\partial a}\right)}^T$ in the appendix. Like previous models, Solving of $\boldsymbol{U}_n^{BUL}=\boldsymbol{0}$ does not have closed form, so we proceed through a numerical optimization methods.
  
\section{Application to the Real data}
In this section, with three real data sets, we will show that the new distributions have a better fit for the bimodal data. \\

\begin{example} \label{realexam1}
The first real data example is a lifetime of $50$ devices put on life test at time $0$, Which is worked out by Aarset (1987). Recently Bakouch et al. (2019) studded this data. Histogram of the data shows that there are two modes. Then we can use a bimodal density function to the data. Bakouch et al. (2019) fitted PLD distribution which has skew and bimodal density function, but ML estimation of the parameters do not stay on the bimodal region of PLD. In this case with fitting BUN, BUSt and BUL, we show that bimodal density function has a better fit. Furthermore, to compare the performance of the new models, we fit BN, PLD, OLLN, BAPN, SN and St distributions to the above-explained data. The results of parameter estimation, AIC and BIC of fitting are reported in table \eqref{tab:1}. According to the AICs and BICs in table \eqref{tab:1} BUN is the most suitable model for data. Furthermore, the Kolmogorov-Smirnov test statistics and P-values are reported in table \eqref{tab:1}, which verifies the goodness of fit under all models for data, as well. Figure \eqref{fig:8} plots the fitted pdfs and the empirical histogram of the data, which demonstrate the flexibility of the BUN distribution.  

\begin{table} [h!]
          \caption{ML inferences under different models for errors for lifetime of $50$ devices data } \label{tab:1}
          \begin{center}
          \resizebox{13cm}{1.5cm}{\begin{tabular}{cccccccccc}
   \hline 
     Model & & & estimates  & & & AIC &    BIC & KS & p-Value \\
\hline
$ \text{BUN}(\mu, \sigma, k) $ & 42.6800& 12.7632 & 2.3325 & & &467.78 & 473.52& 0.0728 & 0.59  \\ 
$\text{BUSt}(\mu, \sigma, k, a, \nu)$ & 42.5099 & 12.6359& 29.8722&  0.2152& 52.1864 & 471.92& 481.48 & 0.0771 & 0.55  \\ 
$\text{BUL}(\mu, \sigma, k, a)$ & 42.0166 & 3.5504 & 0.3404 & -0.4422 & & 487.43 &  495.08 & 0.1067 & 0.32  \\ 
$\text{BN}(\mu, \sigma, a, b)$ & 52.4717 & 4.1059 & 0.0221 & 0.0380 & & 481.98 & 489.63 & 0.1097 & 0.30 \\ 
$\text{PLD}(\nu, \rho, \zeta)$ & 0.0246  &  0.1572 & 1710.2743 & & & 474.02 & 479.75 & 0.1510& 0.10 \\
$\text{OLLN}(\alpha, \mu, \sigma)$ & 42.6193 & 5.3097 & 0.0665 & & & 469.19 & 474.93 & 0.0903 & 0.44 \\ 
$\text{BAPN}(\alpha, \beta, \xi, \eta)$ & 6.8879 &  0.0473 & 42.1179 & 22.2656 & & 474.60 & 482.25 &  &   \\
$\text{SN}(\mu, \sigma, \lambda)$ & 0.0991 & 54.4961& 235116 & & & 478.45 & 484.19 & 0.1375 & 0.15 \\ 
$\text{St}(\mu, \sigma, \lambda, \nu)$ & 0.0869 & 54.3947 & 11100.9 & 4689.1& & 480.48 & 488.13 & 0.1373 & 0.15 \\ 
\hline 
      \end{tabular}}
      \end{center}
      \end{table}

 \begin{figure}  [h!]
 \centering
\includegraphics[scale=0.5]{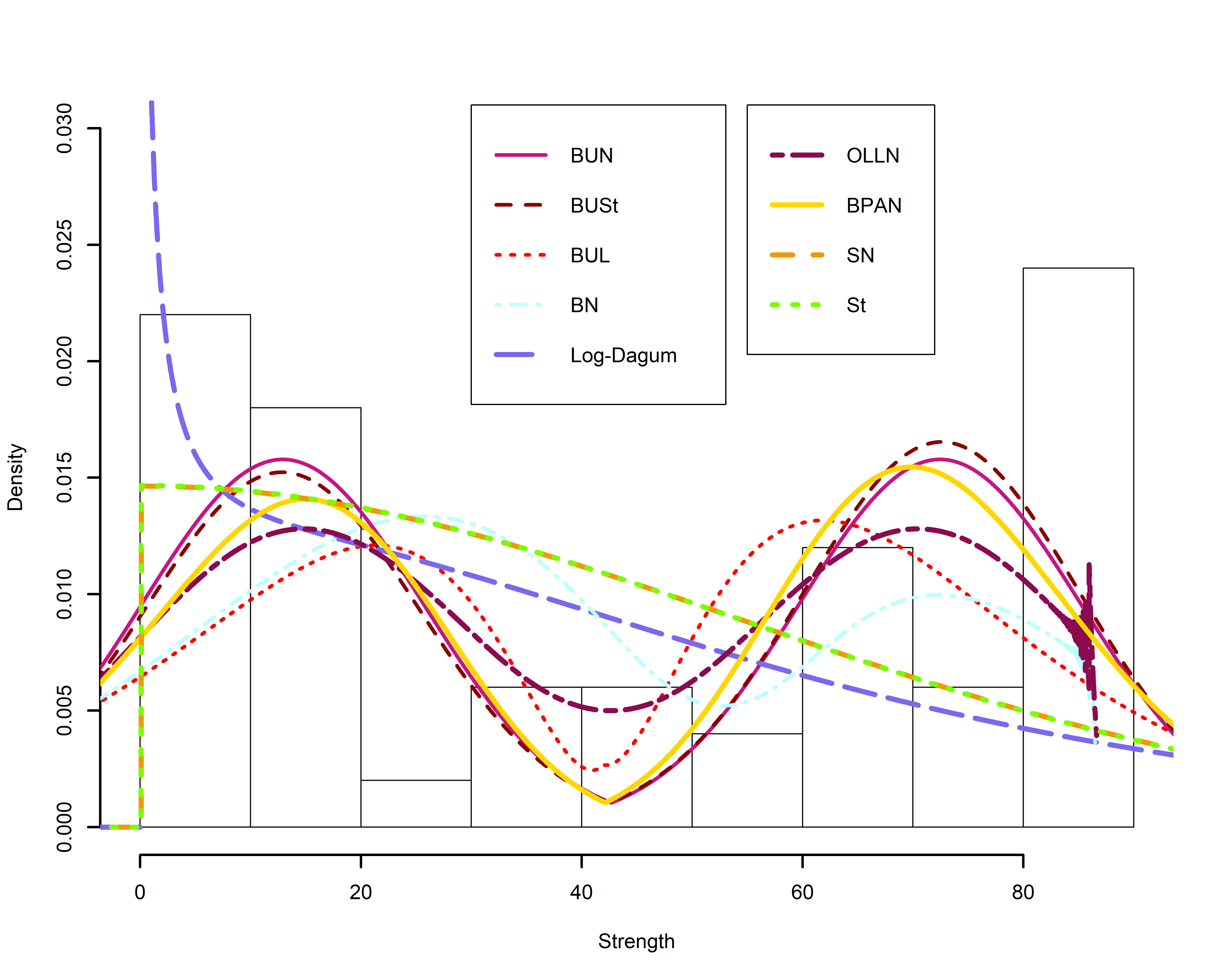}
  \caption{Histograms of the data and the corresponding estimated densities for lifetime of $50$ devices data} \label{fig:8}
  \end{figure} 
\end{example}

\begin{example}
As a second real data example, we studded variable b.weight from the data set considered in Bolfarine et al. (2013). This data set includes 500 observation, and the variable b.weight is the ultrasound weight (fetal weight in grams). These data are available for downloading at \href{url}{http://www.mat.uda.cl/hsalinas/data/weight.rar}. According to the histogram of the data, they are bimodal and symmetric. We fitted the new introduced model and BN, PLD, BAPN and OLLN for data. The results of ML estimation are reported in table \eqref{tab:2}. According to the AICs and BICs in table \eqref{tab:2} BUN distribution is the most flexible model to these bimodal data. The Kolmogorov-Smirnov test statistics and P-values, which are reported in table \eqref{tab:2} shows that all models have been fitted correctly to the data. Fitted pdfs, and the empirical histogram of the data is shown in figure \eqref{fig:9} that the flexibility of the BUN distribution is visible.   

\begin{table} [h!]
          \caption{ ML inferences under different models for errors for ultrasound weight data } \label{tab:2}
          \begin{center}
          \resizebox{13cm}{1.5cm}{\begin{tabular}{cccccccccc}
   \hline 
     Model & & & estimates & & & AIC &    BIC & KS & p-Value \\
\hline
$ \text{BUN}(\mu, \sigma, k) $ &  3213.3166 & 498.4040  &  1.2402 &  &  & 8087.44 & 8100.08 & 0.0209 &  0.64  \\  
$\text{BUSt}(\mu, \sigma, k,\nu)$ & 3212.6529 & 505.2531 & 612.6168 & 417.0372 & & 8089.62& 8106.48 & NAN & NA  \\ 
$\text{BUL}(\mu, \sigma, k)$ & 3201.2857 & 100.5165  &  0.3976 & & & 8099.21 & 8111.85 & 0.0149 & 0.8  \\ 
$\text{BN}(\mu, \sigma, a, b)$ & 3506.5995 & 208.7271 & 0.0773 & 0.1329 & & 8117.62 & 8134.48 & 0.0273 & 0.47 \\ 
$\text{PLD}(\nu, \rho, \zeta)$ & 0.0016 & 0.0048 & 1710.3496 & & & 8224.15 & 8236.79 & 0.0385 & 0.23 \\
$\text{OLLN}(\alpha, \mu, \sigma)$ & 3226.2404 & 252.1888 & 0.1894 & & & 8096.70 & 8109.35 & 0.0319 & 0.36 \\
$\text{BAPN}(\alpha, \xi, \eta)$ & 3.7471 & 3209.4570 &  660.6860 & & & 8088.21 & 8100.85 &  &   \\
 \hline 
      \end{tabular}}
      \end{center}
      \end{table}

  \begin{figure}  [h!]
  \centering
 \includegraphics[scale=0.5]{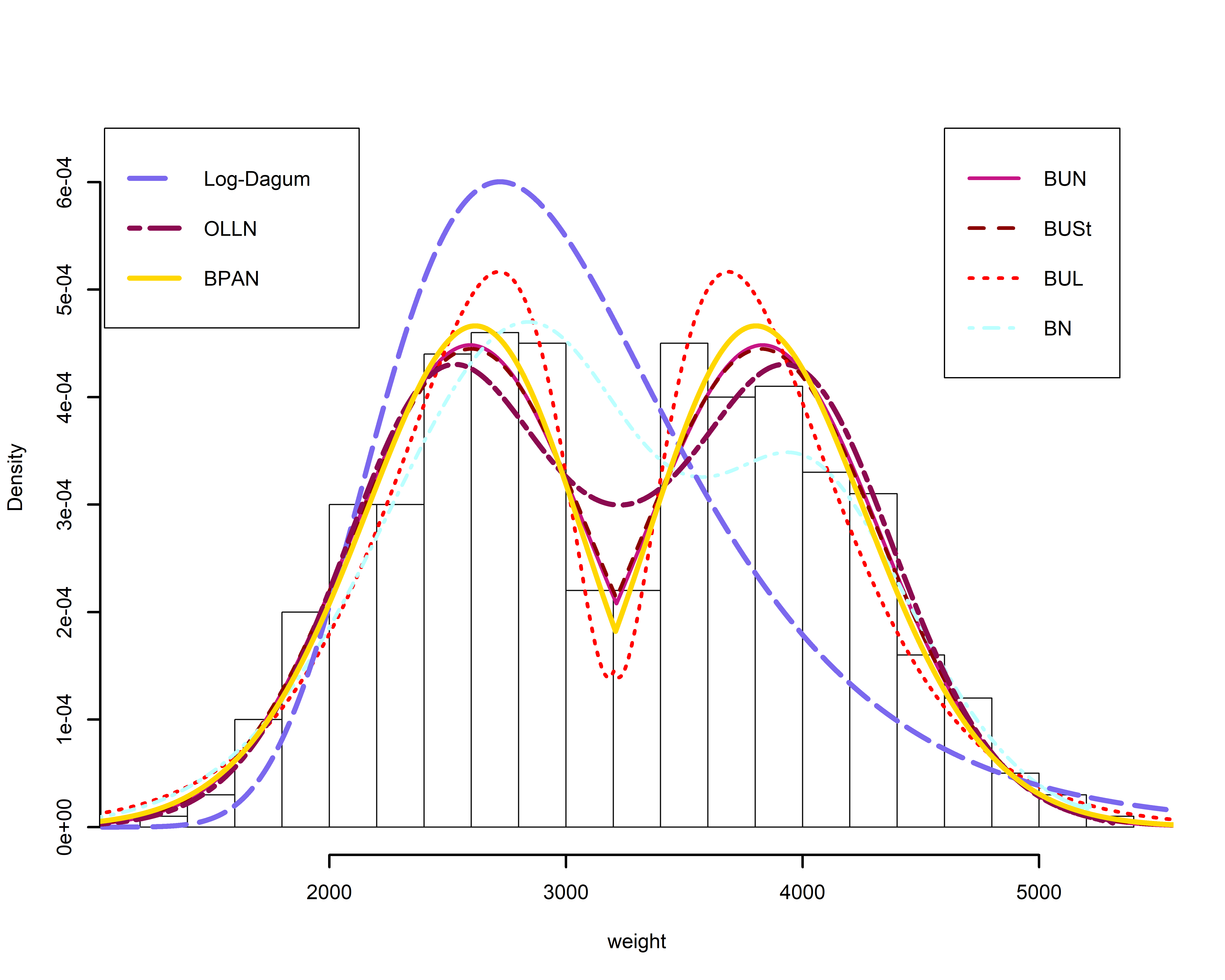}
   \caption{Histograms of the data and the corresponding estimated densities for ultrasound weight data.} \label{fig:9}
   \end{figure}

\end{example}

\begin{example} \label{realexam3}
Third real data example are the lean body mass of Australian athletes. This data have been analyzed by Cook and Weisberg (2009), Asgharzadeh et al. (2013), Sastry and Bhati (2016). The data are given in the appendix.\\
Advantage of BUN, BUSt and BUL is they fit into skewed and bimodal data. To show this fact, We fit new bimodal-unimodal distributions to the lean body mass data, which are skewed data. Also, we fit BN, PLD, OLLN, BAPN, SN and St distributions to the data to compare the new models with chosen old skew distributions. The results of parameter estimation, AIC and BIC are described in table \eqref{tab:3}. According to the AICs and BICs in table \eqref{tab:3} BUL distribution is the best model for the data. The Kolmogorov-Smirnov test statistics and P-values are given in table \eqref{tab:3}. It verifies the fit of all models to the data. Figure \eqref{fig:10} plots the fitted pdfs and the empirical histogram of the data, which demonstrate the flexibility of the BL distribution.  

\begin{table} [h!]
          \caption{ ML inferences under different models for errors for Australian athletes. This data } \label{tab:3}
          \begin{center}
          \resizebox{13cm}{1.5cm}{\begin{tabular}{cccccccccc}
   \hline 
     Model & & & estimates  & & & AIC &    BIC & KS & p-Value \\
\hline
$ \text{BUN}(\mu, \sigma, k, a) $ & 54.6299 & 11.7357 &-1.4901&  0.7996 & & 673.29 & 683.71 & 0.0324 & 0.81  \\ 
$\text{BUSt}(\mu, \sigma, k, a, \nu)$ & 54.6295 & 10.0016 &-12.6934  & 0.8034 & 33.1377 & 675.57 & 688.60 & 0.0416 & 0.71  \\
$\text{BUL}(\mu, \sigma, k, a)$ & 53.4097 &  3.2203 &  1.2239 & -1.5922  & & 666.60 & 677.02 & 0.0461 & 0.65  \\
$\text{BN}(\mu, \sigma, a, b)$ & 65.8970 & 10.2662 &  1.0495 &  4.5749 & & 676.18 & 686.60 & 0.0709 & 0.37  \\
$\text{PLD}(\nu, \rho, \zeta)$ & 0.1340 & 0.0445 & 1710.3349  & & & 702.66 & 710.47 & 0.1076 & 0.10  \\
$\text{OLLN}(\alpha, \mu, \sigma)$ & 55.0707 & 17.3130 &  2.7999 & & & 673.92 & 681.73 & 0.0534 & 0.56  \\
$\text{BAPN}(\alpha, \beta, \xi, \eta)$ &  3.4328 &  1.2655 & 47.4646 &  8.2050 &  & 674.31 & 684.73 & &   \\
$\text{SN}(\mu, \sigma, \lambda)$ & 54.7805 &  6.8832 &  0.0201 & & & 675.73 & 683.54 & 0.0580 & 0.51  \\
$\text{St}(\mu, \sigma, \lambda, \nu)$ & 59.0816 &  7.2329 & -0.9021 &  9.9109 & & 675.48 & 685.90 & 0.0626 & 0.46  \\
 \hline 
      \end{tabular}}
      \end{center}
      \end{table}

 \begin{figure}  [h!]
 \centering
\includegraphics[scale=0.5]{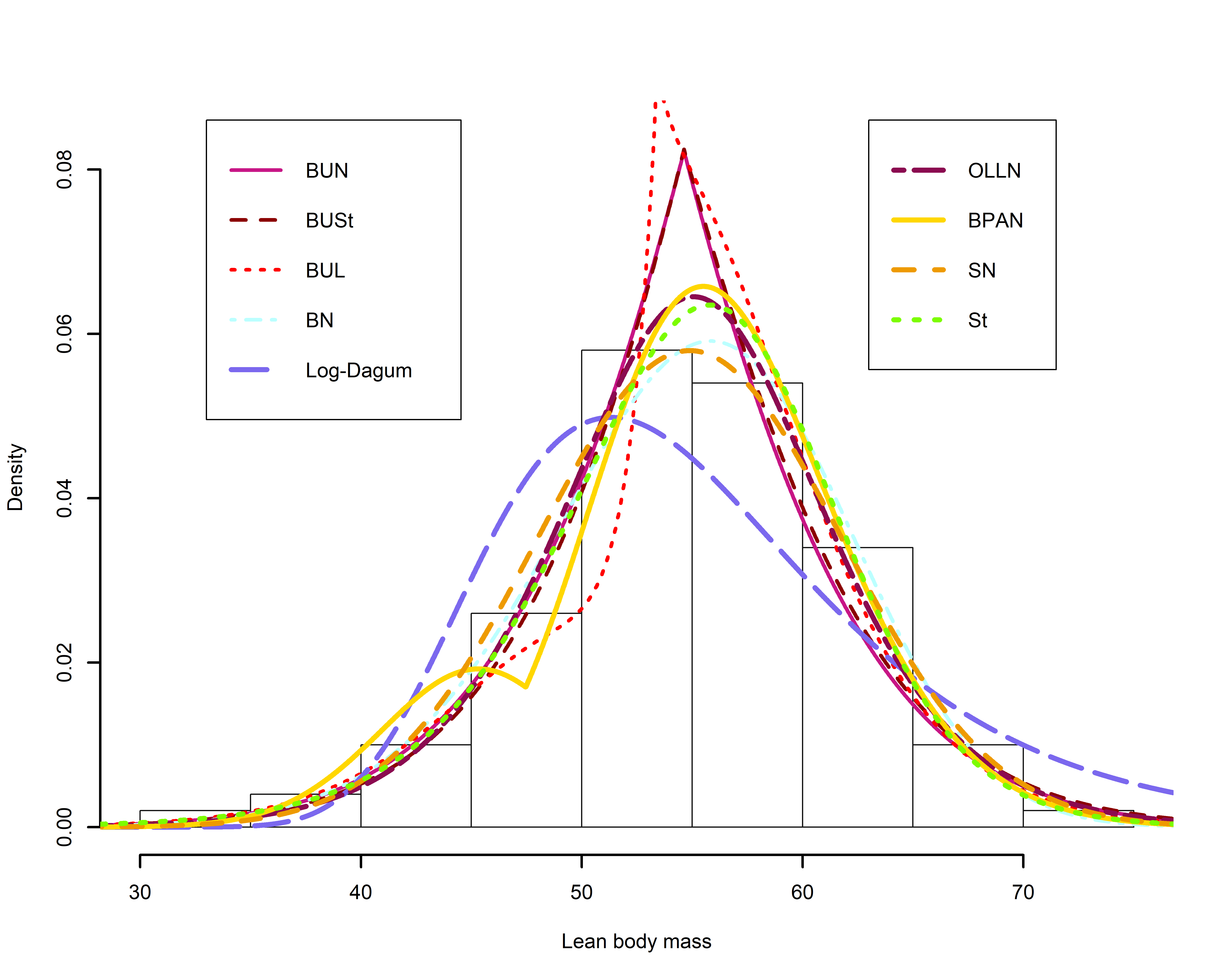}
  \caption{Histograms of the data and the corresponding estimated densities for Australian athletes. This data.} \label{fig:10}
  \end{figure} 
\end{example}

\section{Extensions}  
\subsection{Log-BU distributions}   
If $ X = \log(T) $ follows the BU distributions, then $T$ is said to follow the log-BU distributions. Log-BU densities are adequate models used to describe the lifetime process under fatigue, cure rate proportional hazard and survival regression models. Log-BU extensions of distributions can inherit flexibility and suitability of original densities. For an example of Log-BU generalizations of BU models, we mention Log-BUN as follow. 

The BUN distribution is defined in terms of bimodal extension of Normal on the real line. Then the variate $ T=\exp\left( X\right) $ follow Log-BUN distribution and has the density function as
\begin{eqnarray} 
 f\left(t\right)&=& c_{\sigma, k, a} \exp \left(k\left|\frac{\log(t)-\mu }{\sigma }\right|\right) \frac{1}{t\sigma} \phi \left(\frac{\log(t)-\mu - a }{\sigma }\right)   \qquad t > 0 \nonumber
\end{eqnarray} 

\begin{figure} [h!]
\centering
\subfloat[]{
\resizebox*{6cm}{!}{\includegraphics{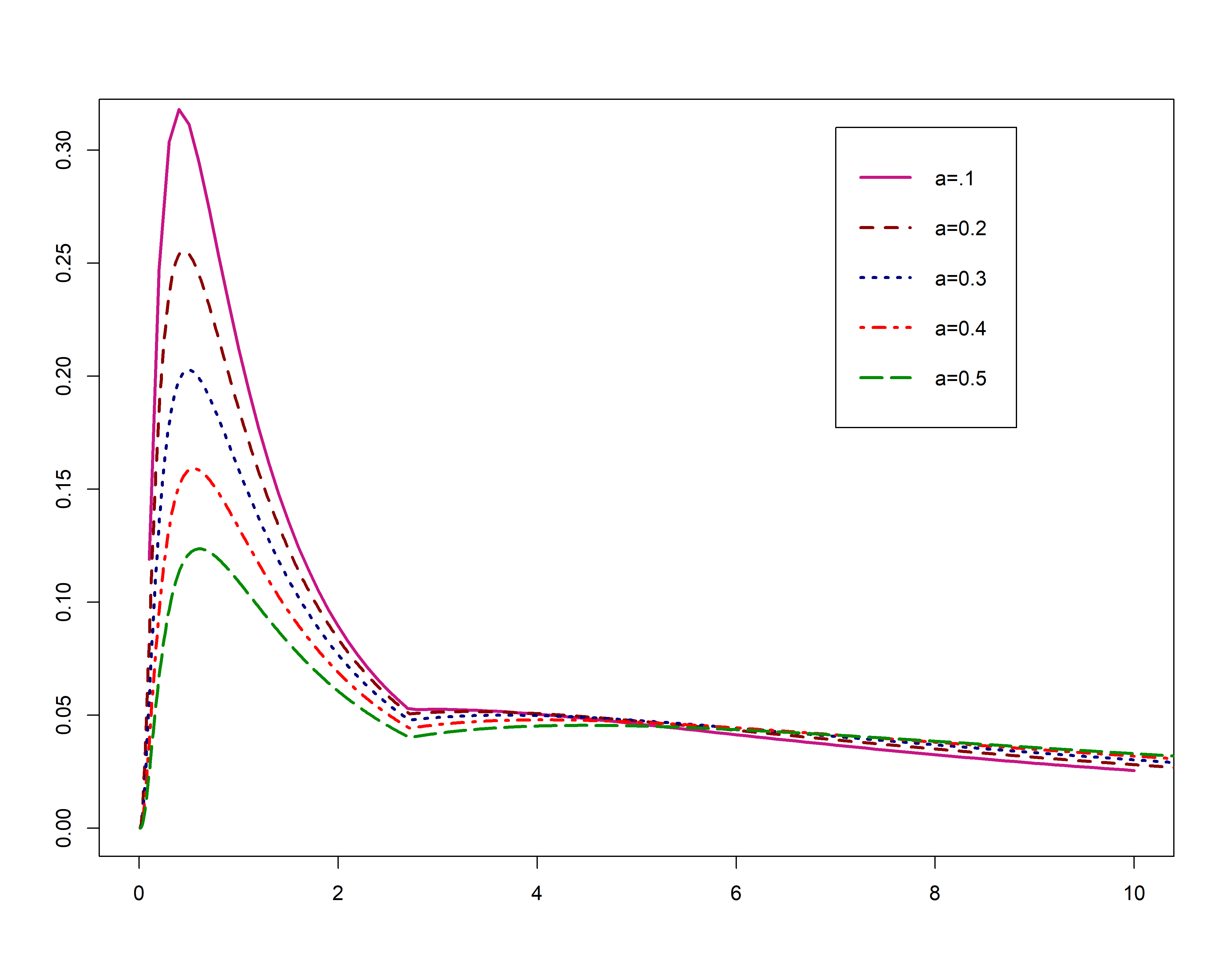}}}\hspace{5pt}
\subfloat[]{
\resizebox*{6cm}{!}{\includegraphics{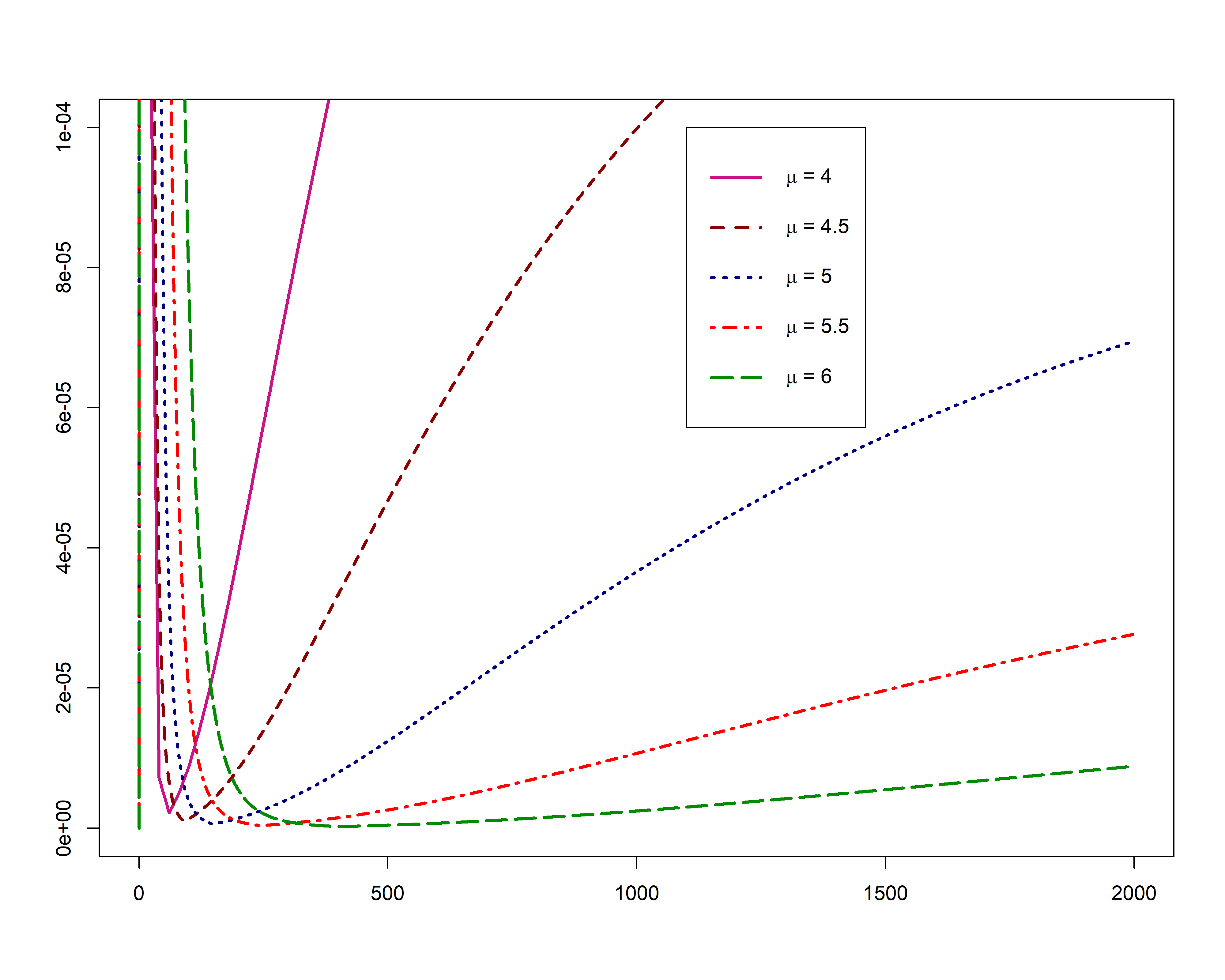}}}
\caption{Shapes of Log-BUN density function for $ \mu=1 $, $\sigma=1$ and $k=1$ part(a) and hazard function for for $\sigma=1$, $k=4$ and $a=0.1$ part (b).} \label{fig:11}
\end{figure}

Where $c_{\sigma, k, a}^{-1}=\exp\left( \frac{ka}{\sigma} +\frac{k^2}{2}\right) \mathrm{\Phi }\left(k+\frac{a}{\sigma}\right)+\exp\left( -\frac{ka}{\sigma} +\frac{k^2}{2}\right)\mathrm{\Phi }\left(k-\frac{a}{\sigma}\right)$. We use the notation $ T \sim \text{Log-BUN}\left(\mu, \sigma, k, a \right) $ when a random variable $ T $ follows the Log-BUN distribution. It is easy to show that $ E(T^r) = M_X(r) = \exp\left( -\dfrac{r\mu}{\sigma}\right)  M_Z\left( \dfrac{r}{\sigma}\right)  $ and $ P\left(T \leq t \right) = F_X(\log(t)) $ where $M_Z(\cdot)$ is the moment generating function of standard BUN random variable and $F_X(\cdot)$ is cdf of BUN distribution. Shapes of Log-BUN density and hazard function for some value of parameters are in figure \eqref{fig:11}.  

\subsection{BU error term regression models}
In many real situations, the error term is skew and (or) bimodal in nature instead of symmetric, which can be modelled by BU distributions. To apply BU distributions in this situation suppose the regression model is $ y_i=\boldsymbol{x}_i' \boldsymbol{\beta}+\epsilon $ for $i=1,2,...,n$, where $\boldsymbol{x}_i$ is covariate vector, $\boldsymbol{\beta}$ is regression coefficients, $y_i$ is response variable and $\epsilon$ is error terms which follow BUN or BUSt or BUL distribution. We can do mean or bimodal regression for this model.

\appendix
\section*{All Technical Details and Proofs}
\subsubsection*{Proof of Theorem \ref{th:2.1}}
$ (i) $ Immediately from symmetric properties of $ g(.) $ and $ w(.) $. \\
$ (ii) $   Beacuse $ f_{X} $ is a density function in $ \mathbb{R} $ then they have at least on mode, Let $ m_1 $ is one of the modes $ f $, without loss of generality let $ m_1 > c $
\begin{eqnarray}
w'(m_1) g(m_1) + w(m_1) g'(m_1) = - w'(2c - m_1) g(2c - m_1) - w(2c - m_1) g'(2c - m_1) = 0 \nonumber
\end{eqnarray}
then the second mode is $ m_2 = 2c - m_1 $. \\ 
$(iii)$ From $(ii)$ we have $ m_1 + m_2 = m_1 + 2c - m_1 = 2c $. 
\subsubsection*{Proof of Theorem \ref{th:2.4}}
Because 
 \begin{center}
 $ b < c$, $w'(c) > 0$
 \end{center}
  and 
\begin{equation}
E [w(X)] f_X'(c) = w'(c) g(c) + w(c) g'(c) = w'(c) g(c)  > 0
\nonumber
\end{equation}
Then $ f_X $, in $ (c,\infty) $ has a mode and let denoted by $ m_1 $. if we find a point in $ (-\infty , b) $ that $ f'_{X} < 0 $ then proof is completed. we know that
\begin{center}
$ 2b - m_1 < 2c - m_1 < b < c  $ 
 \end{center}
and we need find sign $ f_X $ at point $ 2c - m_1 $. We have
\begin{eqnarray}
 0 & = & w'(m_1) g(m_1) + w(m_1) g'(m_1) \nonumber \\
 &= & - w'(2b - m_1) g(2c - m_1) - w(2b - m_1) g'(2c - m_1) \nonumber \\
&= &\dfrac{w'(2b - m_1)}{w(2b - m_1)} + \dfrac{g'(2c - m_1)}{g(2c - m_1)} \nonumber
\end{eqnarray}
Then 
\begin{eqnarray}
   f'_{X} (2c - m_1) &=& w'(2c - m_1) g(2c - m_1) + w(2c - m_1) g'(2c - m_1)  \nonumber \\
& =  & \dfrac{w'(2c - m_1)}{w(2c - m_1)} + \dfrac{g'(2c - m_1)}{g(2c - m_1)} = \dfrac{w'(2b - m_1)}{w(2b - m_1)} - \dfrac{w'(2c - m_1)}{w(2c - m_1)} < 0. \nonumber
\end{eqnarray}
\subsubsection*{Proof of Theorem \ref{th:2.8}}
We first proof that $ f_X(x) $ are density functions and then the bimodal properties are concludes from theorem \eqref{th:2.1} and fact that $ e ^ {k G \left( |x| \right) } $ and $ G\left( |x| \right) ^ k $ are convex.
\begin{eqnarray}
\int_{-\infty}^{\infty} e ^ {k G \left( |x| \right) } g(x) dx  & = &   \int_{-\infty}^{0} e ^ {k G \left( -x \right) } g(x) dx + \int_{0}^{\infty} e ^ {k G \left( |x| \right) } g(x) dx \nonumber \\
 & = & \int_{-\infty}^{0} e ^ {k \left( 1 - G \left( x \right) \right)  } g(x) dx + \int_{0}^{\infty} e ^ {k G \left( x \right) } g(x) dx  \nonumber \\
 & = & \int_{0}^{\frac{1}{2}} e ^ {k \left( 1 - u \right)  }  du + \int_{\frac{1}{2}}^{1} e ^ {k u } du = \dfrac{2 \left( e ^ {k} - e ^ {\frac{k}{2}} \right) }{k}.  \nonumber
 \end{eqnarray} 
 and
\begin{eqnarray}
\int_{-\infty}^{\infty} G \left( |x| \right) ^ k g(x) dx  & = &   \int_{-\infty}^{0} G \left( -x \right) ^ k  g(x) dx + \int_{0}^{\infty} G \left( x \right) ^ k  g(x) dx \nonumber \\
 & = & \int_{-\infty}^{0} \left( 1 - G \left(  x \right) \right)  ^ k  g(x) dx + \int_{0}^{\infty} G \left( x \right) ^ k  g(x) dx  \nonumber \\
 & = & \int_{-\infty}^{0} \left( 1 - u \right)  ^ k  du + \int_{0}^{\infty} u ^ k  du = \dfrac{2 \left( 1 - \frac{1}{2^{k+1}} \right) }{k+1}. \nonumber
\end{eqnarray}

\subsubsection*{Proof of Theorem \ref{th:2.10}}
$ (i) \quad  h'(x)g'(h(x)) = 0 \Longrightarrow h'(x) =  0 \Longrightarrow x = d $ \\
 $ (ii) \quad  h'(x)g'(h(x)) = 0 \Longrightarrow h'(x) = 0 $ or $ g'(h(x)) = 0 \Longrightarrow h(x) = k $ and from convexity of $ h(x), h(x) = k $ have two solution. \\
 $(iii)$ from $ (ii) $.
 
\subsubsection*{Proof of Theorem \ref{th:2.15}} 
remember the pdf \eqref{eq:2} and let $ h(x) = |x| $, then
\begin{eqnarray} 
 f_{X}(x) = \dfrac{g\left( |x| \right) }{\int_{-\infty}^{\infty} g\left( |w|  \right) dw }. \nonumber
\end{eqnarray}
we can rewrite as  
\[ f_{X}(x) = \left\{ \begin{array}{ll}
         \dfrac{g\left( x \right) }{\int_{0}^{\infty} g\left( w  \right) dw } & \mbox{if \quad $x \geq 0$};\\
        \dfrac{g\left( -x \right) }{\int_{-\infty}^{0} g\left( -w  \right) dw } & \mbox{if \quad $x < 0$}.\end{array} \right. \] \nonumber
 or 
 \[ f_{X}(x) = \left\{ \begin{array}{ll}
          \dfrac{g\left( x \right) }{1 - G_{X}(0) } & \mbox{if \quad $x \geq 0$};\\
         \dfrac{g\left( -x \right) }{ G_{-X}(0) } & \mbox{if \quad $x < 0$}.\end{array} \right. \] \nonumber
Where $ G_{X} $ and $ G_{-X} $ are cdf of $ X $ and $ -X $, respectively. then we can rewrite $ f_X $ as
\[ f_{X}(x) = \left\{ \begin{array}{ll}
          \dfrac{g\left( x \right) }{1 - G_{X}(0) } & \mbox{if \quad $x \geq 0$};\\
         \dfrac{g\left( -x \right) }{1 - G_{X}(0) } & \mbox{if \quad $x < 0$}.\end{array} \right. \] \nonumber
 or
 \begin{eqnarray} 
  f_{X}(x) = \frac{1}{2}\dfrac{g\left( |x| \right) }{G_{(X-k)} (k) }. \nonumber
\end{eqnarray} 

\subsubsection*{Elements of BUN score vector}
\begin{eqnarray}
\frac{\partial \ell }{\partial \mu} &=& -\frac{k}{\sigma }\sum^n_{i=1}{\text{sign}\left(x_i-\mu\right)} + \frac{1}{\sigma }\sum^n_{i=1}{\left(\frac{x_i-\mu-a}{\sigma }\right)} \nonumber \\
\frac{\partial \ell }{\partial \sigma } &=& -\frac{n}{\sigma } + n \frac{kas}{\sigma^2} -\frac{k}{{\sigma }^2}\sum^n_{i=1}{\left|x_i-\mu\right|} + \frac{1}{\sigma} \sum^n_{i=1}{\left(\frac{x_i-\mu-a}{{\sigma }}\right)^2} \nonumber \\
\frac{\partial \ell }{\partial k} &=& -nk - n\frac{as}{\sigma}-n\rho + \sum^n_{i=1}{\left|\frac{x_i-\mu}{\sigma }\right|} \nonumber \\
\frac{\partial \ell }{\partial a}  &=& - n\frac{ks}{\sigma} +  \frac{1}{\sigma}\sum^n_{i=1}{\left(\frac{x_i-\mu-a}{\sigma }\right)}  \nonumber
\end{eqnarray}
Where 
\begin{eqnarray}
\delta &=&{\mathrm{e}}^{\frac{ka}{\sigma}}\mathrm{\Phi }\left(k+\frac{a}{\sigma}\right)+{\mathrm{e}}^{-\frac{ka}{\sigma}}\mathrm{\Phi }\left(k-\frac{a}{\sigma}\right) \nonumber \\
s&=&\frac{{\mathrm{e}}^{\frac{ka}{\sigma}}\mathrm{\Phi }\left(k+\frac{a}{\sigma}\right)-{\mathrm{e}}^{-\frac{ka}{\sigma}}\mathrm{\Phi }\left(k-\frac{a}{\sigma}\right)}{\delta } \nonumber \\ 
\rho &=&\frac{2{\mathrm{e}}^{\frac{ka}{\sigma}}\phi \left(k+\frac{a}{\sigma}\right)}{\delta } \nonumber 
\end{eqnarray}

\subsubsection*{Elements of BUSt score vector}
\begin{align}
\frac{\partial \ell }{\partial \mu } &=  \sum_{i = 1}^{n} \dfrac{\dfrac{\nu + 1}{\nu} \dfrac{u^{-}_i}{s_{-}}  I\left(x_i \geq \mu \right) }{ 1 + \dfrac{1}{\nu} \left( \dfrac{u^{-}_i}{\sqrt{s_{-}}}\right) ^ 2 } + \sum_{i = 1}^{n} \dfrac{\dfrac{\nu + 1}{\nu} \dfrac{u^{+}_i}{s_{+}}  I\left(x_i < \mu \right)}{ 1 + \dfrac{1}{\nu} \left( \dfrac{u^{+}_i}{\sqrt{s_{+}}}\right) ^ 2 }  \nonumber  
\displaybreak \\
\frac{\partial \ell }{\partial \sigma } &= n \nu \sigma\dfrac{  s_{-} ^ {-\frac{\nu+2}{2}} D^{-}_{\nu} + s_{+} ^ {-\frac{\nu+2}{2}} D^{+}_{\nu}    }{ \delta } 
\nonumber  \\
&-n \sigma \dfrac{\left( a+k\right)  s_{-} ^ {-\frac{\nu+3}{2}} d^{-}_{\nu} +  \left( k-a\right)  s_{+} ^ {-\frac{\nu+3}{2}} d^{+}_{\nu}}{ \delta }  \nonumber \\
& + \dfrac{\nu+1}{\nu} \sum_{i=1}^{n} \dfrac{\dfrac{\sigma}{s_{-} ^ 2} \left( \dfrac{u^{-}_i}{\sqrt{s_{-}}}\right) ^ 2  I\left(x_i \geq \mu \right) }{1 + \dfrac{1}{\nu} \left( \dfrac{u^{-}_i}{\sqrt{s_{-}}}\right) ^ 2} - \left(\nu +1 \right)\dfrac{\sigma}{s_{-}} \sum_{i = 1}^{n} I\left(x_i \geq \mu \right) \nonumber \\
&+ \dfrac{\nu+1}{\nu} \sum_{i=1}^{n} \dfrac{\dfrac{\sigma}{s_{+} ^ 2} \left( \dfrac{u^{+}_i}{\sqrt{s_{+}}}\right) ^ 2  I\left(x_i < \mu \right) }{1 + \dfrac{1}{\nu} \left( \dfrac{u^{+}_i}{\sqrt{s_{-}}}\right) ^ 2}  - \left(\nu +1 \right)\dfrac{\sigma}{s_{+}} \sum_{i = 1}^{n} I\left(x_i < \mu \right)  \nonumber \\
\frac{\partial \ell }{\partial k } &= - n \dfrac{ a s_{-} ^ {-\frac{\nu+2}{2}} D^{-}_{\nu} + s_{-} ^ {-\frac{\nu+1}{2}} \left( \left( a+k\right) \dfrac{a s_{-} ^ {-1}}{\nu} + 1 \right)  d^{-}_{\nu}  }{ \delta } \nonumber \\
&- n \dfrac{ a s_{+} ^ {-\frac{\nu+2}{2}} D^{+}_{\nu} + s_{+} ^ {-\frac{\nu+1}{2}} \left( \left( k-a\right) \dfrac{a s_{+} ^ {-1}}{\nu} - 1 \right)  d^{+}_{\nu}  }{ \delta } \nonumber  \\
& - \dfrac{\nu+1}{\nu} \sum_{i=1}^{n} \dfrac{\left( \dfrac{u^{-}_i}{\sqrt{s_{-}}}\right) \left(\dfrac{-1 + \dfrac{a s_{-} ^ {-1} }{\nu} \left( u^{-}_i\right) }{\sqrt{s_{-}}} \right)  I\left(x_i \geq \mu \right) }{1 + \dfrac{1}{\nu} \left( \dfrac{u^{-}_i}{\sqrt{s_{-}}}\right) ^ 2}  \nonumber \\
& - \dfrac{\nu+1}{\nu} \sum_{i=1}^{n} \dfrac{\left( \dfrac{u^{+}_i}{\sqrt{s_{+}}}\right) \left(\dfrac{1 - \dfrac{a s_{+} ^ {-1} }{\nu} \left( u^{+}_i\right) }{\sqrt{s_{+}}} \right)  I\left(x_i < \mu \right)  }{1 + \dfrac{1}{\nu} \left( \dfrac{u^{+}_i}{\sqrt{s_{+}}}\right) ^ 2}  \nonumber \\
& +  a \dfrac{\nu + 1}{\nu s_{-}} \sum_{i = 1}^{n} I\left(x_i \geq \mu \right) -  a \dfrac{\nu + 1}{\nu s_{+}} \sum_{i = 1}^{n} I\left(x_i < \mu \right) \nonumber  \\
\frac{\partial \ell }{\partial a } &= - n \dfrac{ k s_{-} ^ {-\frac{\nu+2}{2}} D^{-}_{\nu} + s_{-} ^ {-\frac{\nu+1}{2}} \left( \left( a+k\right) \dfrac{k s_{-} ^ {-1}}{\nu} + 1 \right)  d^{-}_{\nu}  }{ \delta } \nonumber \displaybreak \\
&- n \dfrac{ k s_{+} ^ {-\frac{\nu+2}{2}} D^{+}_{\nu} + s_{+} ^ {-\frac{\nu+1}{2}} \left( \left( k-a\right) \dfrac{k s_{+} ^ {-1}}{\nu} - 1 \right)  d^{+}_{\nu}  }{ \delta } \nonumber  \\
& - \dfrac{\nu+1}{\nu} \sum_{i=1}^{n} \dfrac{\left( \dfrac{u^{-}_i}{\sqrt{s_{-}}}\right) \left(\dfrac{-1 + \dfrac{k s_{-} ^ {-1} }{\nu} \left( u^{-}_i\right) }{\sqrt{s_{-}}} \right)  I\left(x_i \geq \mu \right) }{1 + \dfrac{1}{\nu} \left( \dfrac{u^{-}_i}{\sqrt{s_{-}}}\right) ^ 2}  \nonumber \\
& + \dfrac{\nu+1}{\nu} \sum_{i=1}^{n} \dfrac{\left( \dfrac{u^{+}_i}{\sqrt{s_{+}}}\right) \left(\dfrac{1 + \dfrac{k s_{+} ^ {-1} }{\nu} \left( u^{+}_i\right) }{\sqrt{s_{+}}} \right)  I\left(x_i < \mu \right) }{1 + \dfrac{1}{\nu} \left( \dfrac{u^{+}_i}{\sqrt{s_{+}}}\right) ^ 2}  \nonumber \\
& +  k \dfrac{\nu + 1}{\nu s_{-}} \sum_{i = 1}^{n} I\left(x_i \geq \mu \right) -  k \dfrac{\nu + 1}{\nu s_{+}} \sum_{i = 1}^{n} I\left(x_i < \mu \right) \nonumber \\
\frac{\partial \ell }{\partial \nu } &= - n \dfrac{ \left(-\dfrac{\log\left( s_{-}\right) }{2} - \dfrac{ak}{ \nu s_{-} } \right)  s_{-} ^ {-\frac{\nu}{2}} D^{-}_{\nu} +   s_{-} ^ {-\frac{\nu+5}{2}} \dfrac{\left( a+k\right) ak}{\nu^2} d^{-}_{\nu}  }{ \delta } \nonumber \\
&- n \dfrac{ \left(-\dfrac{\log\left( s_{+}\right) }{2} + \dfrac{ak}{ \nu s_{+} } \right)  s_{+} ^ {-\frac{\nu}{2}} D^{+}_{\nu} -   s_{+} ^ {-\frac{\nu+5}{2}} \dfrac{\left( k-a\right) ak}{\nu^2} d^{+}_{\nu}  }{ s_{+} ^ {-\frac{\nu}{2}}   D_{\nu}\left( \dfrac{   a + k }{\sqrt{s_{-}}}\right) + s_{+} ^ {-\frac{\nu}{2}}   D_{\nu}\left( \dfrac{  k - a }{\sqrt{s_{+}}}\right) } \nonumber \\
& -\dfrac{\log\left( s_{-}\right) }{2} \sum_{i = 1}^{n} I\left(x_i \geq \mu \right) - \dfrac{\nu + 1}{\nu^2} \dfrac{ak}{s_{-}} \sum_{i = 1}^{n} I\left(x_i \geq \mu \right) \nonumber \\
& - \dfrac{\log\left( s_{+}\right) }{2} \sum_{i = 1}^{n} I\left(x_i < \mu \right) +  \dfrac{\nu + 1}{\nu^2} \dfrac{ak}{s_{+}} \sum_{i = 1}^{n} I\left(x_i < \mu \right)   \nonumber \\
& + n \left(  \psi\left(\dfrac{\nu + 1}{2} \right) -  \psi\left(\dfrac{\nu}{2} \right) - \dfrac{1}{\nu} \right) \nonumber \\
&  - \dfrac{1}{2} \sum_{i = 1}^{n} \left\{ \log\left( 1 + \dfrac{1}{\nu} \left( \dfrac{u^{-}_i}{\sqrt{s_{-}}}\right) ^ 2 \right)   I\left(x_i \geq \mu \right) \right\} \nonumber \\
& - \dfrac{\nu+1}{2} \sum_{i = 1}^{n} \dfrac{-\dfrac{1}{\nu ^ 2} \left( \dfrac{u^{-}_i}{\sqrt{s_{-}}}\right) ^ 2 + \dfrac{2ak}{\nu^3 s_{-}} \left( \dfrac{u^{-}_i}{\sqrt{s_{-}}}\right) ^ 2 I\left(x_i \geq \mu \right) }{\left( 1 + \dfrac{1}{\nu} \left( \dfrac{u^{-}_i}{\sqrt{s_{-}}}\right) ^ 2 \right)}     \nonumber \\
&  - \dfrac{1}{2} \sum_{i = 1}^{n} \left\{ \log\left( 1 + \dfrac{1}{\nu} \left( \dfrac{u^{-}_i}{\sqrt{s_{+}}}\right) ^ 2 \right) I\left(x_i < \mu \right) \right\}   \nonumber \\
& - \dfrac{\nu+1}{2} \sum_{i = 1}^{n} \dfrac{-\dfrac{1}{\nu ^ 2} \left( \dfrac{u^{-}_i}{\sqrt{s_{+}}}\right) ^ 2 - \dfrac{2ak}{\nu^3 s_{-}} \left( \dfrac{u^{-}_i}{\sqrt{s_{+}}}\right) ^ 2 I\left(x_i < \mu \right) }{\left( 1 + \dfrac{1}{\nu} \left( \dfrac{u^{-}_i}{\sqrt{s_{+}}}\right) ^ 2 \right)}     \nonumber
\end{align}
Where $ \delta = s_{-} ^ {-\frac{\nu}{2}}   D_{\nu}\left( \dfrac{   a + k }{\sqrt{s_{-}}}\right) + s_{+} ^ {-\frac{\nu}{2}}   D_{\nu}\left( \dfrac{  k - a }{\sqrt{s_{+}}}\right) $, $u^{-}_i = x_i - \mu - a - k$, $ u^{+}_i = x_i - \mu - a + k$, $D^{-}_{\nu} = D_{\nu} \left(\dfrac{a+k}{\sqrt{s_{-}}} \right)$, $D^{+}_{\nu} = D_{\nu} \left(\dfrac{k-a}{\sqrt{s_{+}}} \right)$, $d^{-}_{\nu} = d_{\nu} \left(\dfrac{a+k}{\sqrt{s_{-}}} \right)$ and $d^{+}_{\nu} = d_{\nu} \left(\dfrac{k-a}{\sqrt{s_{+}}} \right)$.

\subsubsection*{Elements of BUL score vector}
\begin{eqnarray}
\frac{\partial \ell }{\partial \mu} &=& \frac{k}{\sigma }\sum^n_{i=1}{\text{sign}\left(x_i-\mu\right)} - \frac{2}{\sigma }\sum^n_{i=1}{\dfrac{\left(\frac{x_i-\mu-a}{\sigma }\right)}{1+\left(\frac{x_i-\mu-a}{\sigma }\right)^2}} \nonumber \\
\frac{\partial \ell }{\partial \sigma } &=& -\frac{n}{\sigma } + \frac{2na^2}{\sigma^3\left( 1+\frac{a^2}{\sigma^2}+\frac{2}{k^2} \right)} +\frac{k}{{\sigma }^2}\sum^n_{i=1}{\left|x_i-\mu\right|} - \frac{2}{\sigma }\sum^n_{i=1}{\dfrac{\left(\frac{x_i-\mu-a}{\sigma }\right)^2}{1+\left(\frac{x_i-\mu-a}{\sigma }\right)^2}} \nonumber \\
\frac{\partial \ell }{\partial k} &=& \dfrac{n}{k} + \dfrac{4n}{k^3\left( 1+\frac{a^2}{\sigma^2}+\frac{2}{k^2} \right)} - \sum^n_{i=1}{\left|\frac{x_i-\mu}{\sigma }\right|} \nonumber \\
\frac{\partial \ell }{\partial a}  &=& - \dfrac{2na}{\sigma^2\left( 1+\frac{a^2}{\sigma^2}+\frac{2}{k^2}\right) } - \frac{2}{\sigma }\sum^n_{i=1}{\dfrac{\left(\frac{x_i-\mu-a}{\sigma }\right)}{1+\left(\frac{x_i-\mu-a}{\sigma }\right)^2}} \nonumber
\end{eqnarray}
     
\end{document}